\documentclass[fleqn,usenatbib]{mnras}
\usepackage{newtxtext,newtxmath}

\usepackage[T1]{fontenc}

\DeclareRobustCommand{\VAN}[3]{#2}
\let\VANthebibliography\thebibliography
\def\thebibliography{\DeclareRobustCommand{\VAN}[3]{##3}\VANthebibliography}

\usepackage{graphicx}	
\usepackage{amsmath}	

\usepackage{amssymb}	
\usepackage{color}      
\usepackage{array}
\usepackage[flushleft]{threeparttable}
\usepackage{booktabs,caption,soul}

\newcommand{\Msun}{\,\mathrm{M}_\odot}
\newcommand{\Msunyr}{\,\mathrm{M_\odot\, yr^{-1}}}
\newcommand{\Oabu}{12+\log(\mathrm{O/H})}
\newcommand{\re}{R_e}
\newcommand{\angstrom}{\mbox{\AA}}
\newcommand{\aref}[1]{\hyperref[#1]{Appendix~\ref{#1}}}

\title[MZR and FZR in non-AGN \& AGN-hosts]{The mass-metallicity and fundamental metallicity relations in non-AGN and AGN-host galaxies}

\author[S.-L. Li et al.]{
Song-Lin Li,$^{1,2}$\thanks{E-mail: songlin.li@anu.edu.au}
Kathryn Grasha,$^{1,2}$\thanks{ARC DECRA Fellow}
Mark R. Krumholz,$^{1,2}$
Emily Wisnioski,$^{1,2}$
Ralph S. Sutherland,$^{1}$
\newauthor
\ Lisa J. Kewley,$^{1,2,3}$
Yan-Mei Chen$^{4,5,6}$
and Zefeng Li$^{1,2,7}$
\\
$^{1}$Research School of Astronomy and Astrophysics, Australian National University, Canberra, ACT 2611, Australia\\
$^{2}$ARC Centre of Excellence for All-Sky Astrophysics in 3 Dimensions (ASTRO 3D), Australia\\
$^{3}$Institute for Theory and Computation, Harvard-Smithsonian Center for Astrophysics, Cambridge, MA 02138, USA\\
$^{4}$School of Astronomy and Space Science, Nanjing University, Nanjing, 210023, China\\
$^{5}$Key Laboratory of Modern Astronomy and Astrophysics (Nanjing University), Ministry of Education, Nanjing, 210023, China\\
$^{6}$Collaborative Innovation Center of Modern Astronomy and Space Exploration, Nanjing, 210023, China\\
$^{7}$Centre for Extragalactic Astronomy, Department of Physics, Durham University, South Road, Durham DH1 3LE, UK
}
\date{Accepted XXX. Received YYY; in original form ZZZ}

\pubyear{2023}

\begin{document}
\label{firstpage}
\pagerange{\pageref{firstpage}--\pageref{lastpage}}
\maketitle

\begin{abstract}
Galaxies' stellar masses, gas-phase oxygen abundances (metallicity), and star formation rates (SFRs) obey a series of empirical correlations, most notably the mass-metallicity relation (MZR) and fundamental metallicity relation (FZR), which relates oxygen abundance to a combination of stellar mass and SFR. However, due to the difficulty of measuring oxygen abundances and SFRs in galaxies that host powerful active galactic nuclei (AGN), to date it is unknown to what extent AGN-host galaxies also follow these correlations. In this work, we apply Bayesian methods to the MaNGA integral field spectrographic (IFS) survey that allow us to measure oxygen abundances and SFRs in AGN hosts, and use these measurements to explore how the MZR and FZR differ between galaxies that do and do not host AGN. We find similar MZRs at stellar masses above $10^{10.5} \Msun$, but that at lower stellar masses AGN hosts show up to $\sim 0.2$ dex higher oxygen abundances. The offset in the FZR is significantly smaller, suggesting that the larger deviation in the MZR is a result of AGN-host galaxies having systematically lower SFRs at fixed stellar mass. However, within the AGN-host sample there is little correlation between SFR and oxygen abundance. These findings support a scenario in which an AGN can halt efficient gas accretion, which drives non-AGN host galaxies to both higher SFR and lower oxygen abundance, resulting in the galaxy evolving off the star-forming main sequence (SFMS). As a consequence, as the SFR declines for an individual system its metallicity remains mostly unchanged. 
\end{abstract}




\section{Introduction}

Gas-phase metallicity, one of the most crucial properties of galaxies, reflects the current chemical state of the Universe, and records the history of a galaxy's star formation, gas accretion, merging, and outflows \citep{maiolino2019}. Oxygen is the most easily measured metal in the gas phase due to its high abundance and large number of optical and ultraviolet (UV) emission lines, and therefore oxygen abundance, defined as $\Oabu$, is a common proxy for total gas phase metallicity. In local star-forming (SF) galaxies, there is a tight relationship between oxygen abundance and stellar mass that spans 5.5 orders of magnitude in stellar mass ($10^{6} \sim 10^{11.5} \Msun$) \citep{tremonti2004,lee2006}, which is known as the stellar mass - metallicity relation (MZR). Along this relation, oxygen abundance rises steeply below $10^{10.5} \Msun$, and flattens above this stellar mass. The existence of the MZR is robust against the choice of the oxygen abundance estimators, although its slope, normalisation, and the exact stellar mass at which it flattens are not \citep[e.g.,][]{andrews2013,kewley2008,lian2015,ly2016,perez2013}. 

Multiple mechanisms have been proposed to explain the MZR. Outflows triggered by stellar feedback might be increasingly metal-rich compared to the mean of the interstellar mediums (ISM) in low-mass galaxies with shallow potential wells, leading them to have highly suppressed metal abundances \citep[e.g.,][]{tremonti2004,chisholm2018,Forbes19,sharda2021b}. Stellar feedback might also reduce the star formation efficiency (SFE) of low mass galaxies, and thus reduce their metal yields \citep{brooks2007}. Yet another possibility is that variations in the integrated galactic initial mass function (IMF) lead to suppressed oxygen yields per unit star formation in dwarf galaxies \citep{koppen2007}.

The MZR also exists at higher redshifts, but with a normalization that monotonically declines with redshift. The relation evolves significantly at $z\sim 2$ \citep[e.g.,][]{erb2006,henry2013,zahid2014} and $z\sim 3$ \citep[e.g.,][]{maier2014,maiolino2008,mannucci2009}. The evolution at $z\sim 2\!-\!3$ has been extended to lower stellar masses and to redshifts up to $z\sim 10$ by recent James Webb Space Telescope (JWST) observations \citep[e.g.,][]{he2024,li2023,nakajima2023}. At $z < 1$, the MZR for massive galaxies appears to match the local one, while the MZR for low-mass galaxies is still evolving \citep[][]{savaglio2005,zahid2011,zahid2013}. This is likely a sign of downsizing, as massive galaxies complete their star formation rapidly and quench around $z\sim 1$, when small galaxies continue forming stars.

Although the MZR exhibits a relatively small scatter ($\sim 0.1$ dex), a number of authors have also explored the extent to which that scatter is correlated with other galactic properties, such as H I mass \citep{bothwell2013}. \cite{mannucci2010} and \cite{lara2010} find that adding a secondary dependence on star formation rate (SFR) to the MZR reduces the scatter to $\sim$ 0.05 dex. The SFR is anti-correlated with the oxygen abundance, and \cite{mannucci2010} propose that this anti-correlation occurs because accretion of pristine gas simultaneously dilutes the metals in the ISM while enhancing the SFR. By combining all available data (at the time) in the redshift range $z < 3.5$, \cite{mannucci2010} propose a new redshift-independent relation among oxygen abundance, SFR, and stellar mass  over the range $z < 2.5$, known as the fundamental metalllicity relation (FZR), and recent JWST results suggest that the FZR remains invariant up to at least $z\sim 8$ \citep{nakajima2023}. In this picture the observed redshift evolution of the MZR is simply a result of galaxies of a fixed stellar mass having higher SFR at higher redshift. The FZR is expressed either as a surface in 3D space constructed by oxygen abundance, SFR, and stellar mass; or a relation between oxygen abundance and a parameter combining stellar mass and SFR. A number of later publications have confirmed the existence of this trend \citep[e.g.,][]{belli2013,brisbin2012,henry2013,nakajima2014}, even though different oxygen abundance calibrators give rise to different normalizations for it \citep[e.g.,][]{andrews2013,cullen2014,maier2014}, a weaker correlation \citep{salim2014}, or even a reversed correlation at the high mass end \citep{yates2012}. A relation like the FZR is also seen in simulations \citep[e.g.,][]{naiman2018,torrey2018}, and is expected from analytic modeling \citep[e.g.,][]{lilly2013,forbes2014,Forbes19}.

Most studies of the FZR and MZR to date focus on galaxies without substantial nuclear activity, since most oxygen abundance calibrators are developed for HII regions and are not reliable when applied to gas ionised by non-stellar radiation sources \citep{kewley2002,kewley2019a}. A few authors have attempted to derive the oxygen abundances of active galactic nuclei (AGN) with direct methods based on the electron temperature \citep[$T_e$,][]{dors2020}, or with strong line ratios \citep[e.g.,][]{storchi1998,castro2017,carvalho2020,dors2021}. By using ultraviolet (UV) strong line ratios, \cite{matsuoka2018} and \cite{dors2019} find that Seyfert galaxies at $z > 1.6$ obey an MZR that is similar to that of non-AGN hosts at the same redshift. By contrast, \citet{armah2023} find that the MZR in local Seyfert galaxies has a lower normalisation than in non-AGN galaxies. However, the oxygen abundance calibrations for Seyfert galaxies used in these studies are based on pure AGN ionising spectra, and thus cannot be applied to galaxies where stellar sources contribute significant ionisation, meaning that comparisons between the AGN- and non-AGN galaxies rely on oxygen abundances derived from different indicators. Since there are well-known systematic differences between oxygen abundance indicators \citep[e.g.,][]{kewley2008}, this makes comparison of the MZR for the two different sub-populations challenging. In addition, even galaxies classified as Seyfert have some contribution to their ionising luminosity from star formation, complicating measurements of oxygen abundance even within the Seyfert sub-sample.
 
 This suggests that a more promising approach to comparing the oxygen abundances of galaxies that do and do not host AGN is to disentangle light from AGN- and SF-driven ionisation, enabling application of a uniform set of line diagnostics with their relative contribution to emission lines being considered. One way to perform this separation is to use Bayesian methods to fit the observed emission as a mixture of the two contributions. \cite{thomas2018a} develop the NebulaBayes code for this purpose, thereby allowing oxygen abundance estimation even in galaxies with active nuclei, and \cite{thomas2019} use this code to investigate the MZR for both SDSS Seyfert and non-Seyfert galaxies. They find that Seyfert galaxies also form a MZR but with generally higher oxygen abundance than non-Seyferts of equal mass. While this is an advance, because it is based on single fibre spectroscopy, separation of the AGN and SF contributions, and thus measurement of the oxygen abundance, remains difficult, and complicated by aperture effects that cause the relative contributions of AGN- and SF-driven light to vary systematically with both the effective radius of the galaxy and its redshift. They also do not explore the FZR in AGN-host galaxies.

Integral field spectroscopy (IFS) surveys of nearby galaxies, such as MaNGA \citep[Mapping nearby Galaxies at Apache Point Observatory,][]{bundy2015}, SAMI \citep[Sydney-AAO Multi-object Integral-field spectrograph,][]{bryant2015}, and CALIFA \citep[Calar Alto Legacy Integral Field Area,][]{sanchez2012}, provide a potential solution. They reveal a clear transition \textit{within} AGN-host galaxies from SF to Seyfert regions as diagnosed via the Baldwin, Phillips \& Terlevich (BPT) diagram \citep{baldwin1981}. This transition is called the mixing sequence \citep{davies2014a,davies2014b,davies2016,dagostino2018}, and by decomposing points within galaxies along this sequence into linear combinations between SF-dominated and AGN-dominated emission, one is able to disentangle the contribution to emission-line flux from AGN and SF sources point by point within a galaxy, and to use the line ratios from SF components alone to estimate the oxygen abundance using the same indicators for all spaxels, thereby removing a significant source of uncertainty. This approach also removes the aperture effects that complicate analysis of single-fibre spectra.

This is the approach we adopt in this work: we combine IFS data from MaNGA with a Bayesian method that allows us to decompose the relative contributions of AGN and SF light spaxel-by-spaxel, and thereby to obtain oxygen abundance measurements using a consistent method across the SF-AGN mixing sequence, along with star formation rate measurements with the AGN contribution to H$\alpha$ emission subtracted. We use these data to investigate the MZR and FZR in a sample including both non-AGN and AGN-host galaxies. The structure for the remainder of this paper is as follows: in \autoref{sec:methods} we introduce the sample selection, classification, and the application of NebulaBayes. We show our analysis of MZR and FZR of both SF and non-SF galaxies in \autoref{sec:results}. Then we discuss and compare our results with previous works, and give a brief summary in \autoref{sec:discussion} and \autoref{sec:conclusion}, respectively.

Throughout the work, we adopt a \cite{chabrier2003} initial mass function (IMF) and a flat WMAP7 cosmology: $H_0=70.4 \,\mathrm{km s^{-1} \,Mpc^{-1}}$, $\Omega_M=0.27$, and $\Omega_{\Lambda}=0.73$.

\section{Methods}
\label{sec:methods}

We use NebulaBayes \citep{thomas2018a} to estimate the spatially-resolved star formation rate (SFR) and oxygen abundance of galaxies drawn from the MaNGA sample \citep{bundy2015}. We start with a brief introduction to MaNGA survey, our sample selection criteria, and our sample classification in \autoref{sec:sample}. We explain our implementation of NebulaBayes in \autoref{sec:NB}. In \autoref{sec:m_sf_abu}, we elucidate how we use the estimation maps output by NebulaBayes to calculate the overall oxygen abundance, SFR, and stellar mass.

\subsection{MaNGA and Sample Selection}
\label{sec:sample}

MaNGA, one of the key projects of Sloan Digital Sky Survey IV \citep[SDSS-IV,][]{blanton2017}, is an IFS survey aiming to gather two-dimensional spectroscopic maps of 10,000 nearby galaxies within $0.01 < z < 0.15$ \citep{wake2017}. The MANGA sample is selected from {\tt v1\_0\_1} of the NASA-Sloan Atlas \citep[NSA,][]{blanton2011} with a flat distribution in $\log(M_*)$ from $10^9$ to $10^{11} \Msun$. The sample covers each target galaxy to at least 1.5 effective radii ($\re$) for all targets \citep{yan2016b}. Consequently, we choose 1 $\re$ as the aperture in this work to remove the aperture effects and ensure most of areas in this aperture can be covered by MaNGA bundles. The emission line and corresponding error maps are extracted from the MaNGA Data Analysis Pipeline (DAP) \citep{belfiore2019, westfall2019}. We take our values of $\re$, axis ratios, and position angles from the elliptical Petrosian fits reported in the MaNGA {\tt DRPAll} catalogue \citep{law2016}, which have proven to be more stable than single-component S{\'e}rsic fits, and are used for MaNGA sample selection \citep{wake2017}. All of the data can be accessed through the portal on the SDSS DR17 website.

\begin{figure*}
    \resizebox{17cm}{!}{\includegraphics{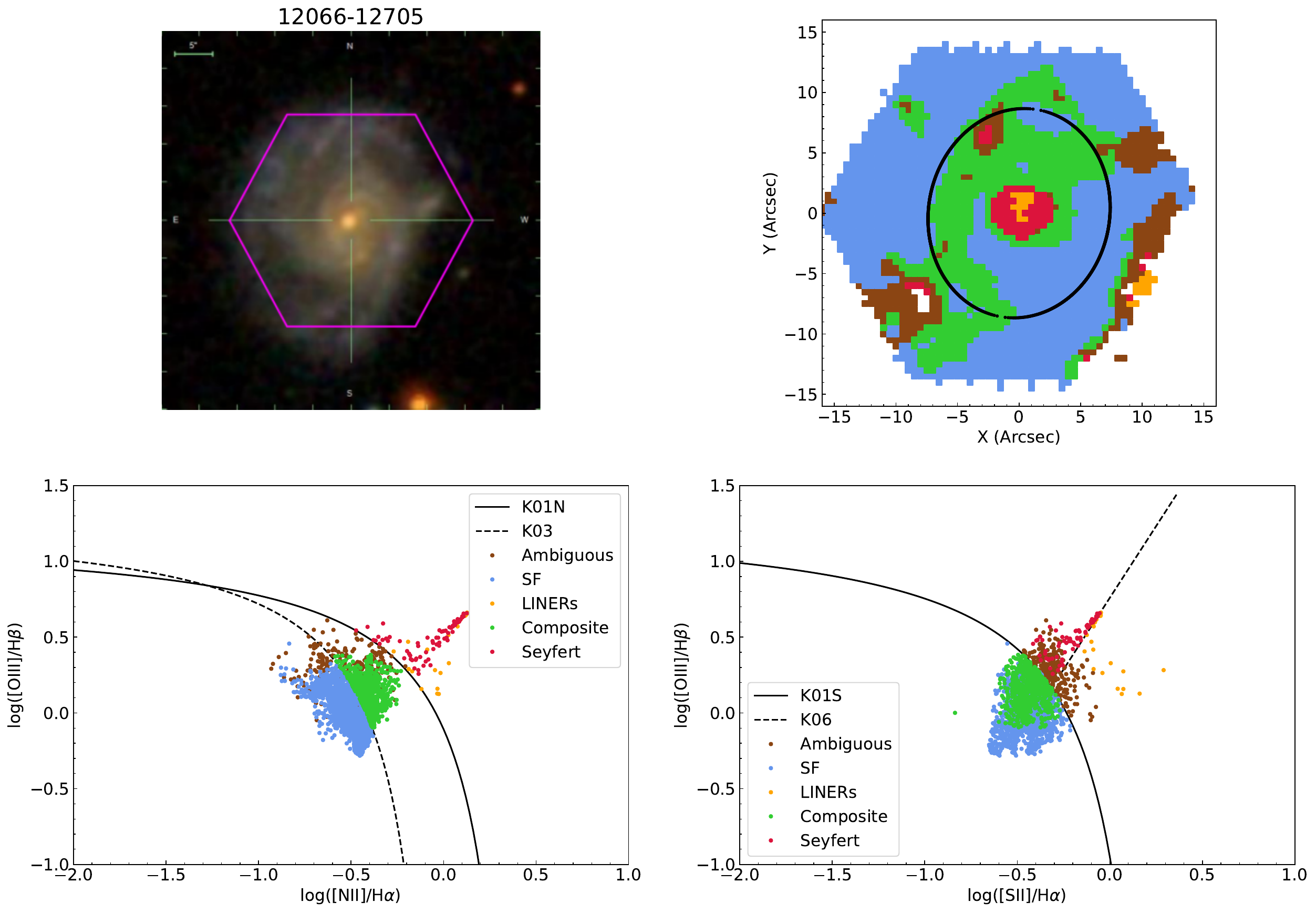}}
    \caption{An example of a galaxy with all five groups of spaxels. \textbf{Top left}: Composite rgb image. The magenta hexagon shows the field of view of MaNGA bundle. The title on the panel is the {\tt PLATEID-IFUID} of this galaxy in MaNGA. \textbf{Top right}: The classification map. The black circle shows 1 $\re$. The color corresponds to the location on the bottom two BPT diagrams (see \autoref{sec:sample}). \textbf{Bottom left}: The $\log(\mathrm{[OIII]/H}\beta$) versus $\log(\mathrm{[NII]/H}\alpha$) BPT diagram. Each dot corresponds to one spaxel in this galaxy. The colors show the classifications in \autoref{sec:sample}. The dashed line is the empirical demarcation line of the pure SF galaxies proposed by \citet{kauffmann2003}. The solid line is the theoretical maximum starburst line from \citet{kewley2001}. \textbf{Bottom right}: The same as the bottom left panel but for the $\log(\mathrm{[OIII]/H}\beta$) versus $\log(\mathrm{[SII]/H}\alpha$) diagram. The solid line is the theoretical maximum starburst line from \citet{kewley2001}. The dashed line is introduced in \citet{kewley2006} to distinguish Seyfert galaxies from LINERs.}
    \label{fig:cls_example}
\end{figure*}

For the galaxy sample we use in this work, we start from the full MaNGA catalogue and firstly remove galaxies with axis ratios smaller than 0.5. This requirement excludes edge-on galaxies to minimise the effects of dust attenuation and to reduce light mixing from regions with various physical conditions. We also remove galaxies with effective radii $R_e < 1\farcs 5$, corresponding to the aperture of SDSS single fiber spectra, and similar to the full width at half maximum (FWHM) of the point spread function (PSF) in $r$-band in MaNGA.

Within each galaxy, we mask spaxels where the signal to noise ratio (SNR) in the H$\beta$ line or H$\alpha$ line is $< 3$, where the provided bitmask of H$\beta$ line or H$\alpha$ line is not 0, or where the reported flux is $\leq 0$ in any of the $\mathrm{[OII]}\,\lambda\lambda 3726,29$, $\mathrm{[OIII]}\,\lambda 5007$, $ \mathrm{[NII]}\,\lambda 6583$, $\mathrm{[SII]}\,\lambda 6716$, or $\mathrm{[SII]}\,\lambda 6731$ lines. These lines are important for calculating priors in NebulaBayes and for classifying spaxels. We show in the \aref{app:snr} that adopting a higher SNR cut does not significantly affect our results. We also mask spaxels with H$\alpha$ equivalent width (EW) $\le 3\,\angstrom$ to remove the contamination from diffuse ionized gas (DIG) ionized by post-asymptotic giant branch (pAGB) stars \citep{belfiore2016,lacerda2018}. This cut is likely sufficient to prevent significant errors in metallicity estimation. Several previous authors have shown that varying the EW cut does not lead to qualitatively different conclusions about galaxy metallicities \citep{thomas2019, baker2023}, and the line ratio that is the most important for driving out oxygen abundance determinations, $\mathrm{[NII]/[OII]}$, is not sensitive to the hardness of the radiation field \citep{kewley2006,kewley2019a}. We further exclude from our sample any galaxy where $> 30\%$ of the spaxels within 1 $\re$ are masked by these conditions, or where visual inspection of the emission spectra from the central $3''$ region shows broad components in either the H$\beta$ or H$\alpha$ lines. The final sample size after applying these criteria is 3,617 galaxies. 

We classify each spaxel based on the original BPT diagram ($\log(\mathrm{[OIII]}\lambda 5007/\mathrm{H}\beta)$ versus $\log(\mathrm{[NII]}\lambda 6583/\mathrm{H}\alpha)$; \citealt{baldwin1981}) and additional $\log(\mathrm{[OIII]}\lambda 5007/\mathrm{H}\beta)$ versus $\log(\mathrm{[SII]}\lambda\lambda 6716,31/\mathrm{H}\alpha)$ diagram from \cite{veilleux1987}. On these diagrams we draw four discrimination lines, following \cite{kewley2006}, and as shown in \autoref{fig:cls_example}:

\begin{itemize}
    \item \textbf{K01N}: The theoretical maximum starburst line on the BPT diagram proposed in \cite{kewley2001}. This is the solid line on the bottom left panel in \autoref{fig:cls_example}. 
    
    \item \textbf{K01S}: The theoretical maximum starburst line on the $\log(\mathrm{[OIII]}/\mathrm{H}\beta) - \log(\mathrm{[SII]}/\mathrm{H}\alpha)$ diagram. This is the solid line on the bottom right panel in \autoref{fig:cls_example}. 
    
    \item \textbf{K03}: An empirical line proposed in \cite{kauffmann2003} to reduce the contamination of the AGN from pure SF galaxies. This is the dashed line on the bottom left panel in \autoref{fig:cls_example}.

    \item \textbf{K06}: Proposed in \cite{kewley2006} to discriminate between Seyfert galaxies and Low-ionisation nuclear emission-line regions (LINERs). This is the dashed straight line on the bottom right panel in \autoref{fig:cls_example}.
\end{itemize}

These four discrimination lines classify every spaxel into five groups:

\begin{itemize}
    \item \textbf{SF}: SF spaxels are those lying below both the \textbf{K01S} and \textbf{K03} lines, which we assume are ionised primarily by star formation. These are marked in blue in \autoref{fig:cls_example}.

    \item \textbf{Composite}: Composite spaxels are those lying below both the \textbf{K01N} and \textbf{K01S} lines, but above the \textbf{K03} line, which likely contain a mixture of SF and AGN-driven ionisation. These are marked in green in \autoref{fig:cls_example}.

    \item \textbf{Seyfert}: Seyfert spaxels are those lying above both the \textbf{K01N} and \textbf{K01S} lines, and above the \textbf{K06} line. These are marked in red in \autoref{fig:cls_example}.

    \item \textbf{LINER}: LINER spaxels are those lying above both the \textbf{K01N} and \textbf{K01S} lines, but below the \textbf{K06} line. These are marked in orange in \autoref{fig:cls_example}.
    
    \item \textbf{Ambiguous}: Ambiguous spaxels are those lying above the \textbf{K01N} line but below the \textbf{K01S} line, or those lying below the \textbf{K01N} line but above the \textbf{K01S} line. These are marked in brown in \autoref{fig:cls_example}.
\end{itemize}

In addition to classifying each spaxel, we also classify galaxies as a whole. Our approach is as follows: we first stack the flux of emission lines required in the BPT diagrams within an aperture $3''$ in diameter, equal to the coverage of an SDSS fiber, and close to the FWHM of the PSF in r-band in MaNGA. All emission lines used for classification have SNRs greater than 3 in the stacked data. We then classify the galaxies based on the ratios the stacked line fluxes in exactly the same way as for individual spaxels.\footnote{For a small fraction of galaxies this approach encounters a problem: in 230 galaxies, more than 70\% spaxels in this aperture are masked due to the $\mathrm{EW(H}\alpha) > 3\, \angstrom$ criterion, indicating the central regions of these galaxies are dominantly ionized by pAGB stars \citep{belfiore2016,lacerda2018}. To handle these galaxies we extend the aperture over which we average until fewer than 70\% of spaxels are masked. The median extended diameter is $6''$.} This procedure yields final counts of 2,551 SF galaxies, 738 Composite galaxies, 84 LINER galaxies, 124 Seyfert galaxies, and 120 Ambiguous galaxies. Hereafter, we will not exhibit relations for Ambiguous group specifically. For much of what follows we will also group the Composite, LINER, Seyfert, and Ambiguous galaxies together into a larger group to which we refer as AGN-host galaxies; this group encompasses all galaxies that show signs of nuclear activity within $3''$, and includes 1,066 galaxies, compared to our 2,551 SF (non-AGN host) galaxies in non-AGN group. Note that our SF category includes all galaxies whose ionisation is powered primarily by star formation regardless of whether they lie on or off the star-forming main sequence (SFMS), defined as an approximate linear relation between $\log \mathrm{SFR}$ and $\log M_*$ for star-forming galaxies. To distinguish this group from the usual definition of SF galaxies in the literature, in the following we will call them as BPT-SF galaxies.

\subsection{NebulaBayes}
\label{sec:NB}

\begin{table} 
 \centering
 \renewcommand{\arraystretch}{1.3}
 \begin{threeparttable}
  \caption{The model grids, NebulaBayes implementation, and the parameter settings of BPT-SF and other spaxels.}
  \vspace{1mm}
  \label{tab:ingredient}

  \begin{tabular}{p{2.5cm}|p{2cm}p{2cm}}
   \hline
   Ingredients & BPT-SF & Others$\mathrm{^a}$ \\
   \hline
   HII-region grids & $\checkmark$ & $\checkmark$ \\
   NLR grids & $\times$ & $\checkmark$ \\
   Line-ratio prior & $\checkmark$ & $\checkmark$ \\
   Full-line likelihood & $\times$ & $\checkmark$ \\
   \hline
   $\log(P/k)$ & Free (20)$\mathrm{^b}$ & Free (20) \\
   $\Oabu$ & Free (160) & Free (160) \\
   $\log U_\mathrm{HII}$ & Free (40) & $-3.25$ \\
   $\log U_\mathrm{NLR}$ & N.A.$\mathrm{^c}$ & Free (40) \\
   $\log(E_\mathrm{peak}/\mathrm{keV})$ & N.A. & Fixed in galaxy$\mathrm{^d}$ \\
   $f_\mathrm{HII}$ & 1 & Free (40) \\
   \hline
  \end{tabular} 
  \begin{tablenotes}
   \item $\mathrm{^a}$Others enclose Composite, LINER, Seyfert, and Ambiguous spaxels. See \autoref{sec:sample} for details.
   \item $\mathrm{^b}$The number in the parentheses is the number of interpolated points when running NebulaBayes.
   \item $\mathrm{^c}$The $\log U_\mathrm{NLR}$ and the $\log(E_\mathrm{peak}/\mathrm{keV})$ are not applicable on SF spaxels.
   \item $\mathrm{^d}$The $\log(E_\mathrm{peak}/\mathrm{keV})$ of non-SF spaxels is fixed for spaxels within one galaxy, but varies among galaxies.
  \end{tablenotes}
 \end{threeparttable}
\end{table}

After selecting and classifying galaxies and their spaxels, the next step in our analysis is using NebulaBayes \citep{thomas2018a} to derive their parameters. NebulaBayes provides a Bayesian posterior parameter estimate by comparing the observed emission-line luminosity with models from photoionisation code MAPPINGS \citep{sutherland2017}. Each model is a linear combination of an HII-region component and a narrow line region (NLR) component, and the free parameters that characterise the model, which NebulaBayes fits, are the gas pressure $\log(P/k)$, the oxygen abundance $\Oabu$, the ionisation parameters in the HII regions and NLRs $\log U_\mathrm{HII}$ and $\log U_\mathrm{NLR}$, respectively, and the peak photon energy $\log(E_\mathrm{peak}/\mathrm{keV})$ of the accretion disk Big Blue Bump component for NLRs. The final free parameter is $f_\mathrm{HII}$, the fractional HII-region contribution to the measured H$\alpha$ flux. Rather than assuming Solar-scaled abundances, we adopt the more accurate Galactic Concordance (GC; \citealt{nicholls2017}) scheme to set the relative abundances of metal atoms as a function of $\Oabu$. We discuss the implications of this choice, in particular with respect to the N/O ratio, in detail in \autoref{sec:caveats}. We provide a full description of how we derive our model spectra and the settings of MAPPINGS in \aref{app:model}. 

In principle we could use NebulaBayes to estimate all the free parameters above in every non-masked spaxel in our sample. However, this is computationally intractable due to the size of the parameter space and the very large number of spaxels in our sample. We therefore using a simplified procedure following \cite{thomas2019} with some modifications, and which generally produces results very close to those produced by a full fit in significantly less computational time. Here we only summarise the procedure, and refer readers to \citeauthor{thomas2019} for a full justification. For readers' convenience we also summarise the procedure in \autoref{tab:ingredient}.

\subsubsection{Procedure for BPT-SF Spaxels}

For spaxels classified as BPT-SF (see \autoref{sec:sample} for details), we assume that only HII regions contribute to the emission lines, so $f_\mathrm{HII}$ is fixed to unity and $\log U_\mathrm{NLR}$ and $\log(E_\mathrm{peak}/\mathrm{keV})$ become irrelevant, leaving $\log(P/k)$, $\Oabu$ and $\log U_\mathrm{HII}$ as the only parameters to be fit. This reduction in the dimensionality of the parameter space greatly accelerates fitting.

We fit these parameters using the ``line-ratio prior'' feature of NebulaBayes. In this mode, rather than attempting to fit all the lines to which we have access simultaneously, we instead run NebulaBayes three times, but each time only for a single line ratio that is particularly sensitive to a single parameter; we use ${\mathrm{[NII]}\,\lambda 6583 / \mathrm{[OII]}\,\lambda \lambda 3726,3729}$
for $\Oabu$ \citep{kewley2002}, $\mathrm{[OIII]}\,\lambda 5007 / \mathrm{[OII]}\,\lambda \lambda 3726,3729$ for $\log U_\mathrm{HII}$ \citep{kewley2002,kobulnicky2004}, and $\mathrm{[SII]}\,\lambda 6716 / \mathrm{[SII]}\,\lambda 6731$ for $\log(P/k)$ \citep{kewley2019b}. The result is three sets of posterior likelihoods for each point in our model grids, but for each set the likelihood varies significantly only in a single parameter, e.g., running NebulaBayes on only the line ratio $\mathrm{[NII]/[OII]}$ returns a posterior likelihood for every combination of $\Oabu, \log U_\mathrm{HII}, \mathrm{and} \log(P/k)$, but in practice the likelihood depends mostly on the first of these three coordinates. For the purposes of this calculation, we follow NebulaBayes's standard treatment of uncertainties, whereby the uncertainty on a given emission line is taken to be the measurement uncertainty combined with a contribution from a 35\% systematic uncertainty in the model flux -- equation 3 of \citet{thomas2018a} for details. We then multiply the three sets of posteriors together, weighting [NII]/[OII] three times as much as the other line ratios since we are most interested in metallicity, to obtain the final posterior probabilities for each model.

\subsubsection{Procedure for Other Spaxels}

For other spaxels including Composite, LINER, Seyfert, and Ambiguous (\autoref{sec:sample}), we use the full model including both HII-region and NLR contributions. However, to reduce the dimensionality of the space and thus the computational time, we fix $\log U_\mathrm{HII} = -3.25$, which is the value in our model grid closest to the median value of $-3.28$ measured in our BPT-SF spaxels, and is the same as the value adopted by \cite{thomas2019}. In addition, since all spaxels within one galaxy are ionised by the same AGN, they should share the same $\log(E_\mathrm{peak}/\mathrm{keV})$. To account for this, in galaxies classified as BPT-SF, Composite, and Ambiguous, where the AGN contribution is relatively small, we fix $\log(E_\mathrm{peak}/\mathrm{keV}) = -1.4$, which is the value in our model closest to  \citeauthor{thomas2019}'s recommendation $-1.35$, and which they found provided good fits for their sample. For Seyfert and LINER galaxies, where the AGN contribution is stronger, we wish to use a more general procedure where we do not assume a fixed value of $\log(E_\mathrm{peak}/\mathrm{keV})$; we therefore determine $\log(E_\mathrm{peak}/\mathrm{keV})$ for these galaxies by selecting the Seyfert and LINER spaxels within central $3''$ of these galaxies and running NebulaBayes with $\log(E_\mathrm{peak}/\mathrm{keV})$ as a free parameter; we then adopt the median value of these spaxels as the value for all other spaxels in the same galaxy. With these simplifications, the final four free parameters are $\log(P/k)$, $\Oabu$, $\log U_\mathrm{NLR}$, and $f_\mathrm{HII}$.

We run NebulaBayes three times. The first two are using ``line-ratio prior'' feature, the same as SF spaxels above, with $\mathrm{[NII]/[OII]}$ and $\mathrm{[SII]}\,\lambda 6716 / \mathrm{[SII]}\,\lambda 6731$. The final run is using the ``full-line likelihood'' feature. Instead of single line ratio, we input 12 available emission lines from MaNGA: $\mathrm{[OII]}\,\lambda\lambda 3726,29$, $\mathrm{[NeIII]}\,\lambda 3869$, H$\beta$,  $\mathrm{[OIII]}\,\lambda 5007$, $\mathrm{HeI}\,\lambda 5876$, $\mathrm{[OI]}\,\lambda 6300$, H$\alpha$, $\mathrm{[NII]}\,\lambda 6583$, $\mathrm{[SII]}\,\lambda 6716$, and $\mathrm{[SII]}\,\lambda 6731$, to obtain the likelihood. These consist of most of the strong optical lines and high-excitation lines typical for Seyfert galaxies.\footnote{For a small fraction of spaxels some of these lines are not well-fit by the MaNGA pipeline, usually due to low signal to noise, and these lines do not have reported fluxes and errors in the MaNGA DAP. To handle these lines, we assign an error of $10^{-16} ~\mathrm{erg\,s^{-1}cm^{-2}\angstrom^{-1}spaxel^{-1}}$, which is $\approx 100-500$ times larger than the typical error for a MaNGA line, and a flux of $5\times 10^{-18} ~\mathrm{erg\,s^{-1}cm^{-2}\angstrom^{-1}spaxel^{-1}}$ (5\% of the error). This assignment ensures that the line carries essentially no weight in our calculation of the likelihood, while still satisfying the formal requirement that the line be assigned a positive flux and error.} We multiply the three sets of posterior together to obtain the final posterior, and determine the value of parameters. Quantitatively, \citet{thomas2019} shows that the typical difference in derived oxygen abundance from different procedures between the BPT-SF galaxies and Seyfert galaxies is $\lesssim 0.03$ dex, much smaller than the results we will obtain below.

\subsubsection{Dereddening}

For the purposes of both oxygen abundance inference and measuring SFRs from the H$\alpha$ emission line, it is important to deredden the observations. We do so using the expression provided in the appendix of \cite{vogt2013}. In the context of NebulaBayes, the deredding procedure is to take the $\mathrm{H}\alpha/\mathrm{H}\beta$ ratio at each model grid point as the intrinsic ratio, and then deredden the remaining lines using the ratio of the observed to intrinsic $\mathrm{H}\alpha/\mathrm{H}\beta$ as input to \citeauthor{vogt2013}'s expression. Upon completion, NebulaBayes returns the H$\alpha$ and H$\beta$ fluxes of the single best-fitting model. We use the ratio of these model-predicted lines together with the observed $\mathrm{H}\alpha/\mathrm{H}\beta$ ratio as input to the \citeauthor{vogt2013} expression to correct the observed H$\alpha$ flux. We use this reddening-corrected value below whenever we wish estimate the SFR or related quantities.

\subsubsection{Sample NebulaBayes Output}

\begin{figure}
{\includegraphics[width=\columnwidth]{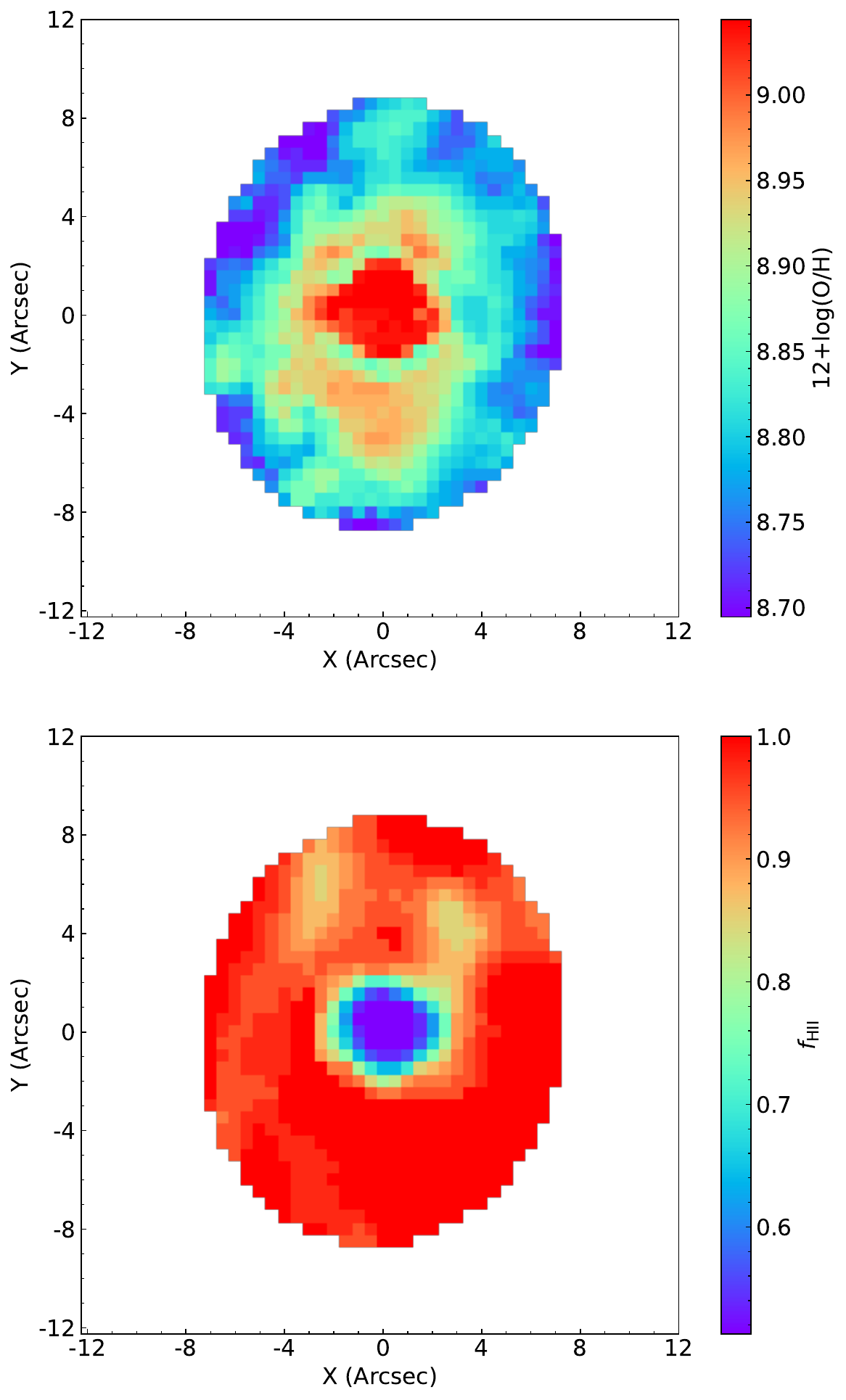}}
    \caption{An example of two estimation maps from NebulaBayes we use in this work. The target is the same as \autoref{fig:cls_example}. The top panel is the oxygen abundance map, and the bottom panel is the $f_\mathrm{HII}$ map.}
    \label{fig:est_map}
\end{figure}

\autoref{fig:est_map} shows two example output maps from NebulaBayes; the top panel shows the median oxygen abundance we obtain in each spaxel, and the bottom shows our median estimate of $f_\mathrm{HII}$. The target shown is the same as in \autoref{fig:cls_example}. We apply NebulaBayes only to spaxels within 1 $\re$, which is why the maps are truncated. The top panel shows that we recover a clear negative oxygen abundance gradient, which is consistent with previous work \citep[e.g.][]{belfiore2017,sanchezm2018}. In contrast, the $f_\mathrm{HII}$ map in the bottom panel shows a positive gradient with decreasing HII-region contribution toward the central region, consistent with the increasingly Seyfert-like line ratios shown in \autoref{fig:cls_example}.

\subsection{The Stellar Mass, SFR, and Oxygen Abundance}
\label{sec:m_sf_abu}

\begin{figure}
{\includegraphics[width=\columnwidth]{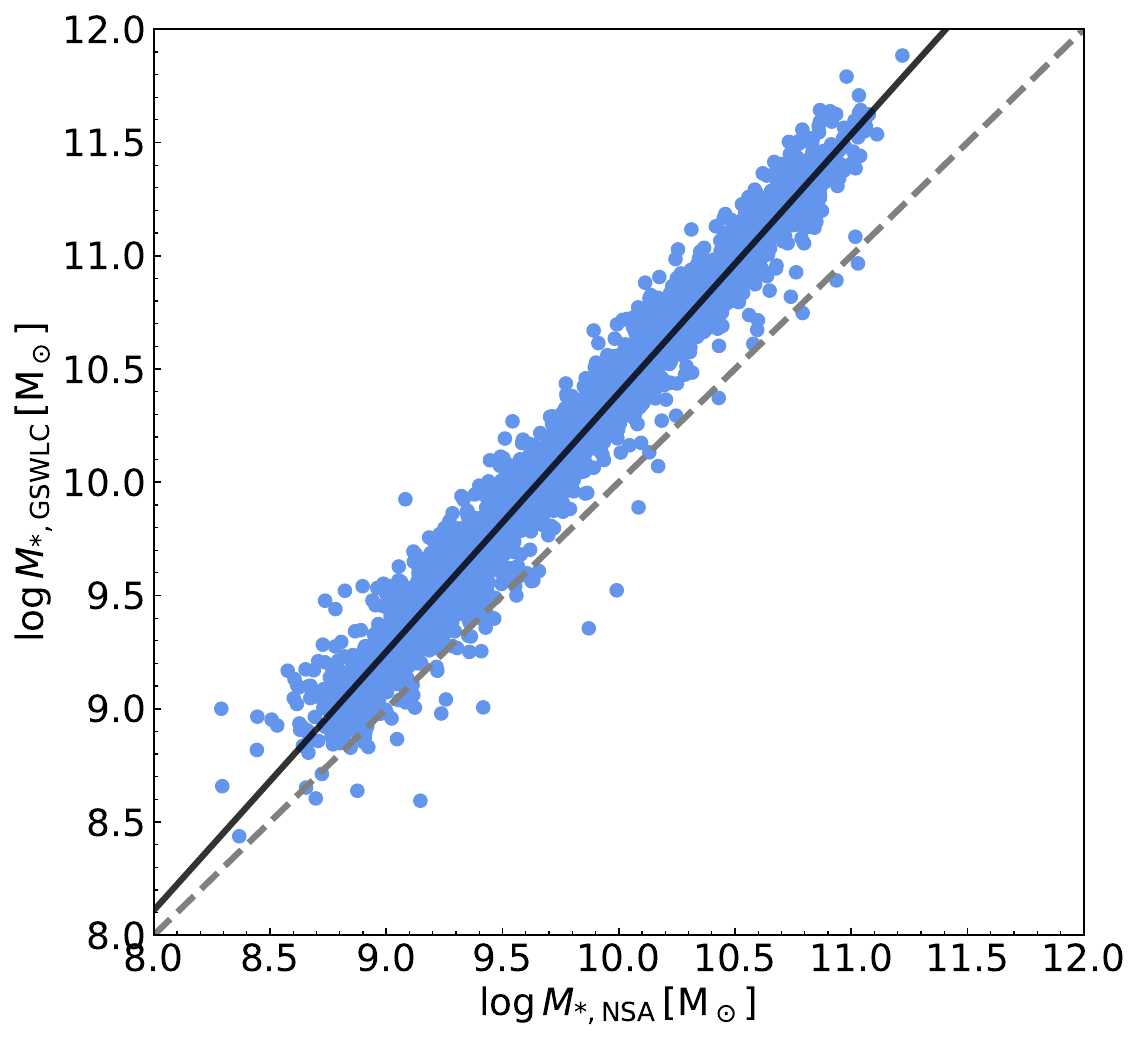}}
    \caption{The relation between $\log M_*$ estimated from the GSWLC and NSA catalogues for the 3,048 galaxies in our sample that appear in both. The grey dashed line is the 1 to 1 relation, and the black line is a linear fit to the data. We use this linear fit to estimate $\log M_*$ for galaxies that are not in the GSWLC.}
    \label{fig:mass_corr}
\end{figure}

\begin{figure}
{\includegraphics[width=\columnwidth]{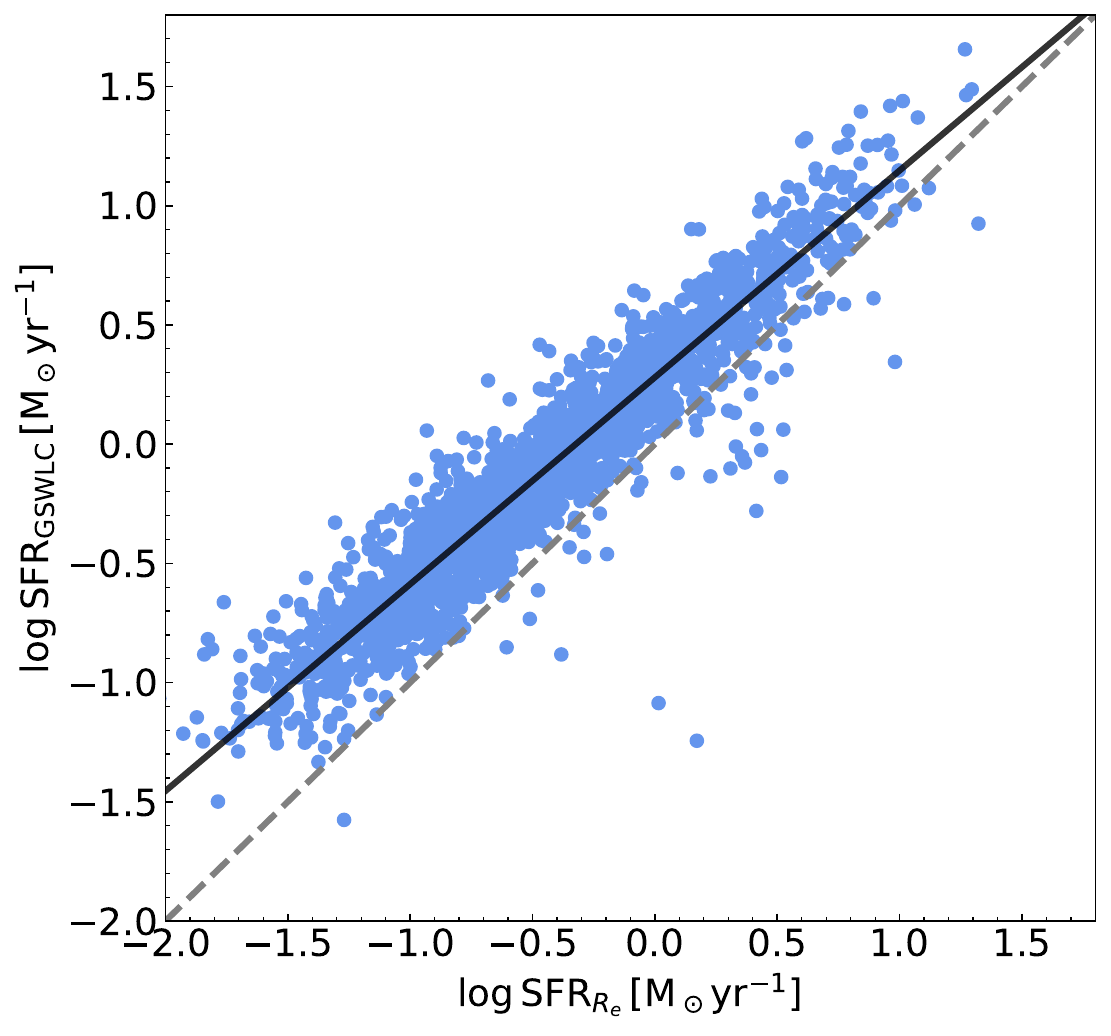}}
    \caption{The relation between $\log \mathrm{SFR}$ from the GSWLC and the $\log \mathrm{SFR}$ within 1 $\re$ of galactic centre estimated in this work (\autoref{equ:sfr_re}) for the 2,163 BPT-SF galaxies in our sample that also appear in the GSWLC. The grey dashed line is the 1 to 1 relation, and the black line is a linear fit to the data. We use this fit to make aperture corrections from 1 $\re$ to whole galaxies to obtain $\log \mathrm{SFR}$.}
    \label{fig:sfr_corr}
\end{figure}

In this section, we discuss how we obtain galaxies' total stellar mass, SFR, and oxygen abundance.

\subsubsection{Stellar Masses}

We take the majority of our values of $\log M_*$ from the GALEX-SDSS-WISE Legacy Catalog \citep[GSWLC;][]{salim2016,salim2018}. This catalogue derives stellar mass and SFR by spectral energy distribution (SED) fitting combining ultra-violet (UV) photometry from the Galaxy Evolution Explorer (GALEX), optical photometry from SDSS, and mid-infrared (MIR) photometry from the Wide-field Infrared Survey Explorer (WISE). However, 569 out of 3,617 galaxies in our sample are not included in the GSWLC. For these galaxies, we instead take $\log M_*$ from the NSA catalogue (which provides stellar masses based on SDSS and GALEX photometry by k-correction from \citet{blanton2007}) and apply a correction to ensure consistency between NSA and GSWLC. Our correction procedure is as follows: in \autoref{fig:mass_corr} we show a comparison of estimated stellar masses for the 3048 galaxies in our sample that appear in both the GSWLC and NSA catalogues. The figure shows that the two stellar mass estimates form a tight relation; a linear fit gives
\begin{equation}
\log M_\mathrm{*,GSWLC}=1.14\log M_\mathrm{*,NSA}-1.03.
\end{equation}
We use this relation to estimate $\log M_*$ for galaxies not in the GSWLC. The origin of the discrepancy between the masses in the two catalogues likely comes from two sources: \citet{blanton2007} and \citet{salim2016} show that up to $\sim 0.2$ dex of difference in estimated stellar mass can result from the dependence of the SED on the star formation history, with the effect being most prominent at high mass and sSFR. The remaining difference may result from differences in their SDSS photometry choices: GSWLC uses {\tt modelMag}, where the magnitude is estimated by assuming a parametric profile, while the NSA catalogue uses {\tt elPetroMag}, where the magnitude is extracted within the elliptical Petrosian radius.

We note that GSWLC does not include an AGN contribution in its fits to the SED, and one might worry that this biases the stellar mass estimates for AGN relative to non-AGN host galaxies. To mitigate this effect, we exclude Seyfert 1 AGN, which have prominent contributions in the UV/optical bands, from our sample by eliminating galaxies that show broad-line features in either the H$\alpha$ or H$\beta$ emission lines of their central spectra (\autoref{sec:sample}). For those AGN that are not removed by this cut, \citet{bellstedt2020} note that the AGN emission is expected to dominate primarily in the MIR bands. However, due to the uncertainties in the photometry of these bands, and the tendency of SED fits including MIR data to produce large residuals, they provide only weak constraints on the stellar mass. As a further cross-check against bias, we compare stellar masses taken from the GSWLC-1 \citep{salim2016} and GSWLC-2 \citep{salim2018} catalogues. GSWLC-1 fits the SED using only UV/optical photometry, while GSWLC-2 (which is used in this work) adds one more MIR band (22 $\mu$m); therefore if MIR emission from AGN produces a substantial bias to our stellar mass estimates, this should show up as a systematic difference in the stellar masses provided by the two catalogues. However, when we compare the stellar masses, we find a 1$\sigma$ scatter of only $\sim 0.02$ dex about the 1-to-1 relation, with no significant difference in the scatter between AGN host and non-AGN host galaxies. This strongly suggests that excess MIR emission from AGN does not significantly influence our stellar mass estimates.

\subsubsection{Star Formation Rates}

Obtaining SFRs for our sample is somewhat more involved than obtaining stellar masses. We cannot simply take values from existing catalogues because, while the GSWLC provides SFR estimates, the NSA does not. Nor do we wish to estimate SFRs solely from our H$\alpha$ maps, because our maps only go out to $1\re$, and thus may miss significant amounts of star formation at larger radii. Instead, our strategy is to combine these two possible approaches: we first measure the SFR within $1\re$ of galaxy centre directly from our data, and then use the difference between that estimate and SFRs taken from GSWLC to develop a model for aperture correction that we can apply to all of our measurements.

In detail, the procedure is as follows. We first compute the star formation rate in the inner $1\re$ of each galaxy, $\mathrm{SFR}_{\re}$, from the sum of dust- and $f_\mathrm{HII}$-corrected H$\alpha$ luminosities of spaxels within 1 $\re$ of the galactic centre,
\begin{equation}
\mathrm{SFR}_{\re}=\left[\sum_{i} L_{\mathrm{H}\alpha,i} f_{\mathrm{HII},i}\right]\times \frac{7.9\times10^{-42}\,\mathrm{M}_\odot\mbox{ yr}^{-1}/(\mbox{erg s}^{-1})}{10^{0.25}},
\label{equ:sfr_re}
\end{equation}
where $L_{\mathrm{H}\alpha,i}$ is the dereddened H$\alpha$ luminosity of the $i$th spaxel and $f_{\mathrm{HII},i}$ is the fraction of the H$\alpha$ emission we attribute to HII regions (and thus to stellar sources) rather than AGN. 
The numerical conversion factor from H$\alpha$ luminosity to SFR is taken from \cite{kennicutt1998b}, with the additional factor of $10^{0.25}$ in the denominator for changing from a \citet{salpeter1955} IMF to a \citet{chabrier2003} IMF, which we use in this work \citep{bernardi2010}.

Next, in \autoref{fig:sfr_corr} we show the relation between $\log \mathrm{SFR}_{\re}$ and the global SFR estimate provided by the GSWLC, $\log \mathrm{SFR_{GSWLC}}$, for the 2163 BPT-SF galaxies in our sample that appear in the GSWLC; we use only BPT-SF galaxies for this purpose because our AGN-host galaxies (\autoref{sec:sample}) generally have small SFR and are off the SFMS, and the SFR from GSWLC-master catalog is not recommended for galaxies not on the SFMS; that said, we find our AGN-host galaxies generally follow the same trend but with larger scatter. The figure shows that the two SFR estimates form a tight relationship with an overall scatter of 0.15 dex. A linear fit gives
\begin{equation}
\log \mathrm{SFR_{GSWLC}}=0.87\log \mathrm{SFR}_{\re}+0.28.
\label{equ:sfr_whole}
\end{equation}

Our procedure to estimate SFRs is therefore to measure $\mathrm{SFR}_{\re}$ for all galaxies, and then compute the total SFR from \autoref{equ:sfr_whole}. To estimate errors in the SFR, we propagate the scatter of the relation to the error in $\log \mathrm{SFR}$. For consistency, we apply this procedure even to BPT-SF galaxies that are included in the GSWLC (i.e., to galaxies where we could in principle simply take $\log \mathrm{SFR}$ directly from the catalogue); however, we have tested the effects of instead using $\log \mathrm{SFR}$ values taken directly from the GSWLC where they are available, and we find that doing so does not change our main conclusions.

\subsubsection{Oxygen Abundance}

\begin{figure}
{\includegraphics[width=\columnwidth]{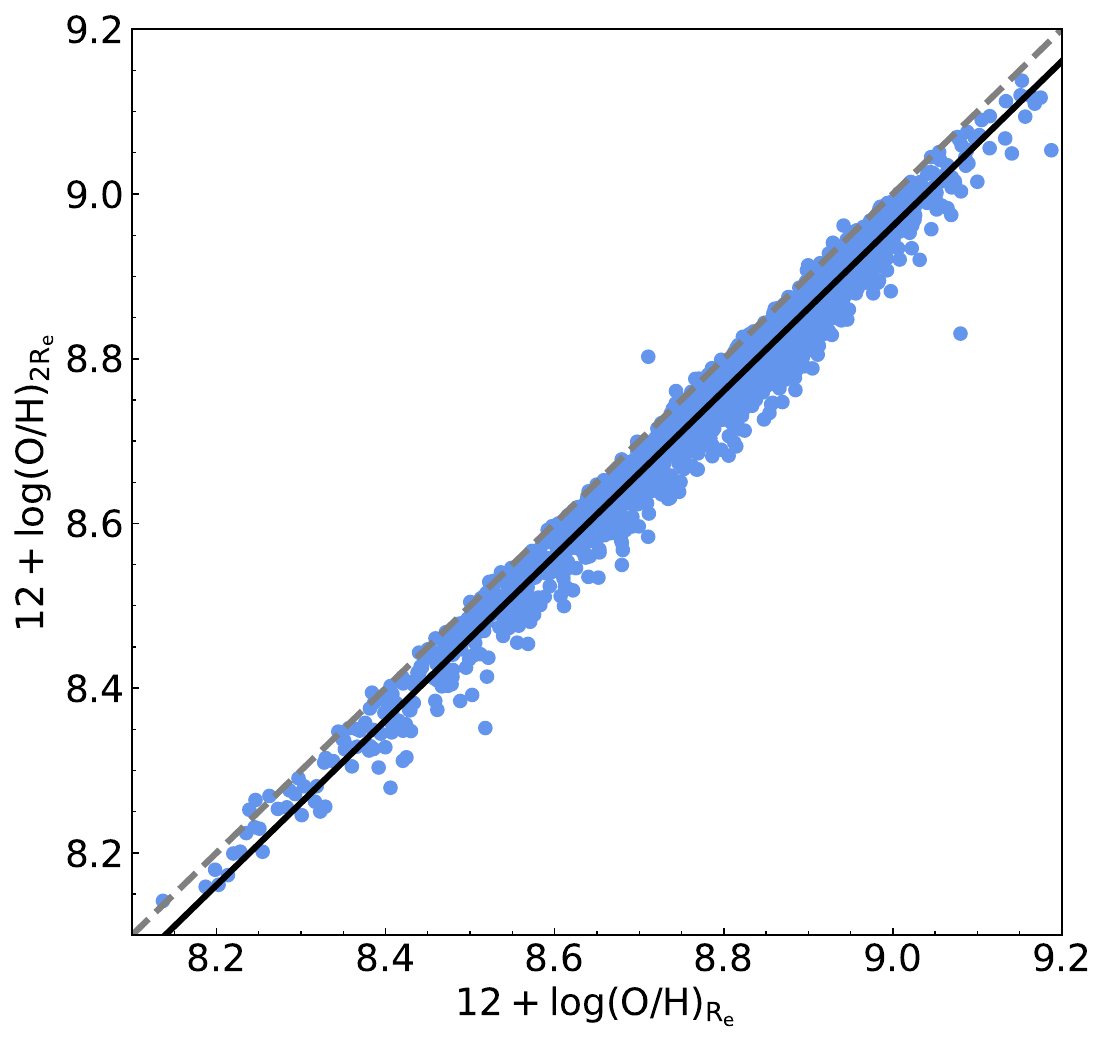}}
    \caption{The oxygen abundance within 2 $\re$ versus the oxygen abundance within 1 $\re$ relation for 2,080 galaxies. The dashed grey line is the 1 to 1 line, and the black solid line is the linear fitting of the relation. The slope of the fitting is 1.00. The oxygen abundance within 2 $\re$ are generally lower by 0.04 dex.}
    \label{fig:justify}
\end{figure}

To obtain the oxygen abundance, we also start by estimating $\Oabu_{\re}$, the SFR-weighted mean oxygen abundance within $1\,\re$, from our NebularBayes maps. We define this quantity as
\begin{equation}
\Oabu_{\re} \equiv 12+\log\left(\frac{\sum_{i} F_i(\mathrm{H}\alpha)_\mathrm{dered} f_{\mathrm{HII},i} (\mathrm{O/H})_i}{\sum_{i} F_i(\mathrm{H}\alpha)_\mathrm{dered} f_{\mathrm{HII},i})}\right),
 \label{equ:abu}
\end{equation}
where the sums run over all non-masked spaxels within 1 $\re$ of galactic centre. We drop the highest and lowest 1\% of the oxygen abundance measurements from this sum to make the average more robust against outliers and spaxels that approach the edges of the range over which the diagnostics are valid and calibrated. In the expression above, $F_i(\mathrm{H}\alpha)_\mathrm{dered}$ is the dust-corrected H$\alpha$ flux in the $i$th spaxel, $f_{\mathrm{HII},i}$ is the corresponding fractional contribution of HII-regions to the H$\alpha$ luminosity obtained from NebulaBayes, and $(\mathrm{O/H})_i$ is the linear oxygen abundance at the $i$th spaxel. The $f_\mathrm{HII}$ factor here ensures that we are weighting by the H$\alpha$ flux from HII regions only, excluding contributions from the NLRs in galaxies with significant AGN emission.

We note that our oxygen abundance measurement is an SFR-weighted mean rather than a mass-weighted mean. This is closer to what is returned by whole-galaxy integrated light measurements, and thus is a more appropriate choice for comparing to the earlier spatially-unresolved data sets from which the MZR and FZR were originally measured. We further note that a number of analytic models also now directly predict SFR-weighted metallicity precisely for the purpose of comparing to observations \citep[e.g.,][]{sharda2021a,sharda2021b}.

It has been proposed that the oxygen abundance at 1 $\re$ agrees with the average abundance throughout the galaxy \citep[e.g.,][]{sanchez2013,sanchez2019}. To verify that this is the case for our data set, we repeat the full analysis pipeline described in \autoref{sec:sample} and \autoref{sec:NB}, but extending the aperture from 1 $\re$ to 2 $\re$. Using this larger aperture results in a final selection of 2,080 out of 3,617 galaxies, with the remaining $\approx 1,550$ removed because, with the larger aperture, they fail to meet our condition that at most $30\%$ of the spaxels within the aperture are masked. For the 2,080 galaxies that remain, we again calculate the average oxygen abundance from \autoref{equ:abu}, simply using all pixels within 2 $\re$ instead of within 1 $\re$ as in our original sample. \autoref{fig:justify} shows the relation between the oxygen abundance within 1 $\re$ and within 2 $\re$ we obtain from this procedure. A linear fit shows that when we double the aperture, the overall oxygen abundance is reduced by 0.04 dex, while the slope of the relationship is 1.00. Thus we conclude that extending the aperture to 2 $\re$ only leads to an overall offset by 0.04 dex but with no bias, and thus does not change the shape of relations such as the MZR or FZR. Given that 0.04 dex is smaller than any plausible estimate of the systematic uncertainties in oxygen abundance calibrations, we simply use $\Oabu_{\re}$ to represent the oxygen abundance of the whole galaxy throughout the remainder of this work.

\section{Results}
\label{sec:results}

In \autoref{sec:mzr} we first examine the MZR and FZR in non-AGN (BPT-SF) galaxies in order to demonstrate that our analysis of these galaxies is consistent with previous work, and to establish a methodology for application and a baseline for comparison to the AGN-host galaxies. In \autoref{sec:other_gal} we extend our analysis to examine the MZR of AGN-host galaxies and compare to the non-AGN sample, and in \autoref{sec:fzr} we carry out a similar comparison with respect to the FZR.

\subsection{The MZR and FZR in Non-AGN Galaxies}
\label{sec:mzr}

\begin{figure}
    {\includegraphics[width=\columnwidth]{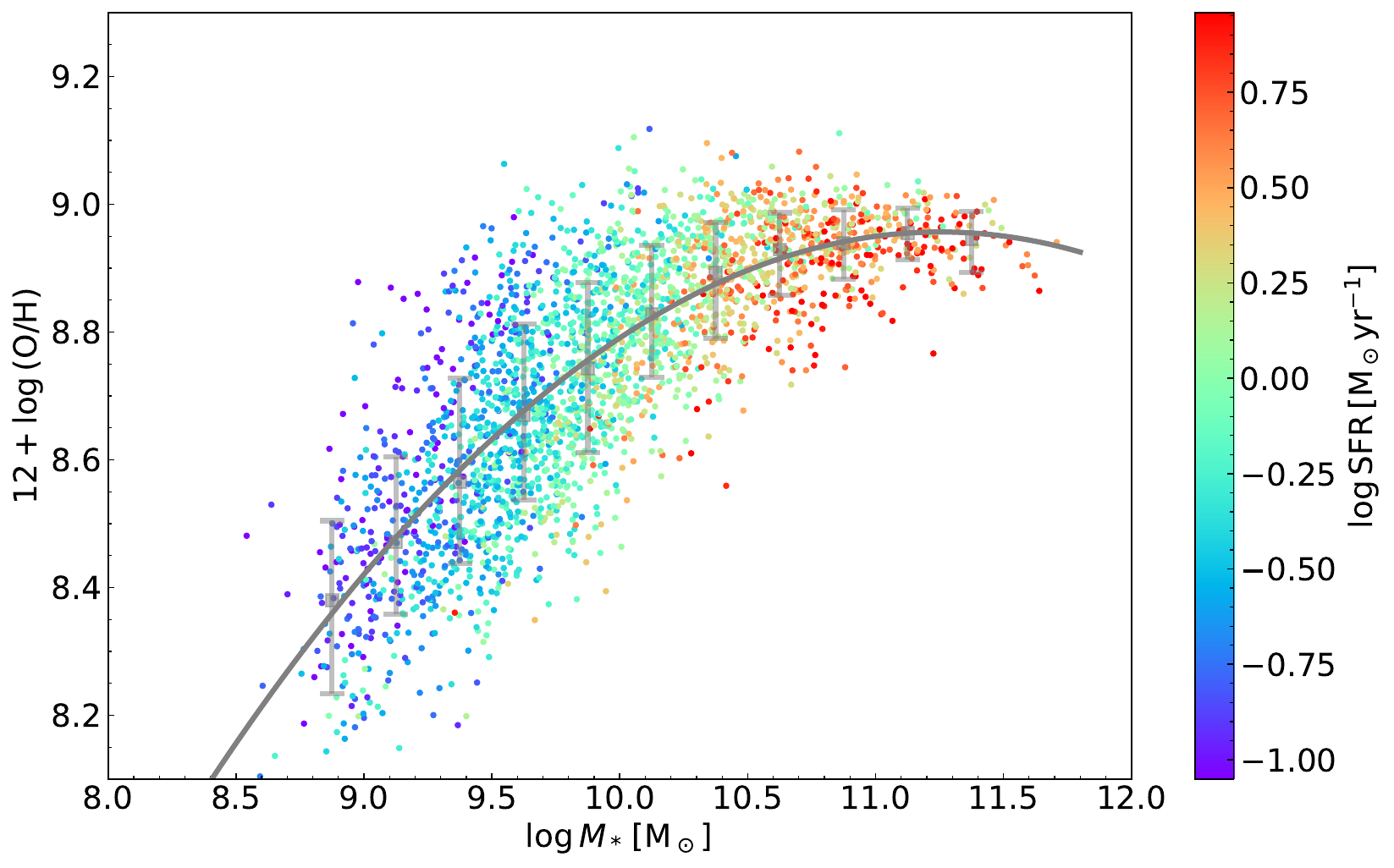}}
    \caption{$\Oabu$ versus $\log M_*$ for BPT-SF (non-AGN) galaxies. Each point represents a single galaxy from our sample, and points are colored by $\log \mathrm{SFR}$. The grey squares and error bars show the median and 1$\sigma$ level scatters of oxygen abundance in bins of stellar mass, and the grey line shows a second-order polynomial function fit to these data (\autoref{equ:mzr}). Note that isocolour sets of points are tilted relative to vertical, indicating a correlation between SFR and oxygen abundance at fixed stellar mass. }
    \label{fig:mzr_sf}
\end{figure}

\autoref{fig:mzr_sf} shows the MZR of BPT-SF (non-AGN) galaxies. Consistent with earlier work, we find a tight relation with a scatter of $\approx 0.11$ dex, with the oxygen abundance increasing with stellar mass until the relation flattens above $\approx 10^{10.5}\rm \Msun$. The scatter is roughly mass-independent and similar to the overall scatter below $10^{10} \Msun$, but does decease to ~0.05 dex at the highest stellar masses - likely a result of lower number statistics. To measure the 1$\sigma$ scatter as a function of stellar mass quantitatively, we first bin the galaxies by $M_*$. We then randomly select 1,000 realisations of the oxygen abundance for every galaxy from a Gaussian distribution with a standard deviation equal to the error of the oxygen abundance measurement; we determine this error by taking the error on the oxygen abundances of individual spaxels that are output by NebulaBayes and propagating the errors through \autoref{equ:abu} for the galaxy-averaged oxygen abundance. We plot the bin medians and uncertainties that we recover from this procedure as grey squares with error bars in \autoref{fig:mzr_sf}, and this plot confirms that the scatter is indeed smaller at higher stellar mass. It is worth noting that our measurement errors are much smaller than the size of the scatter we find in the MZR, suggesting that scatter we see in \autoref{fig:mzr_sf} is true galaxy-to-galaxy variation in oxygen abundance at fixed stellar mass, as shown by previous authors \citep[e.g.,][]{forbes2014}. We fit the binned average values and uncertainties we recover with a second-order polynomial function, and find a best fit
\begin{equation}
\Oabu=-0.106 (\log M_*)^2+2.38 \log M_*-4.41.
\label{equ:mzr}
\end{equation}

\begin{figure*}
    \resizebox{17cm}{!} 
    {\includegraphics{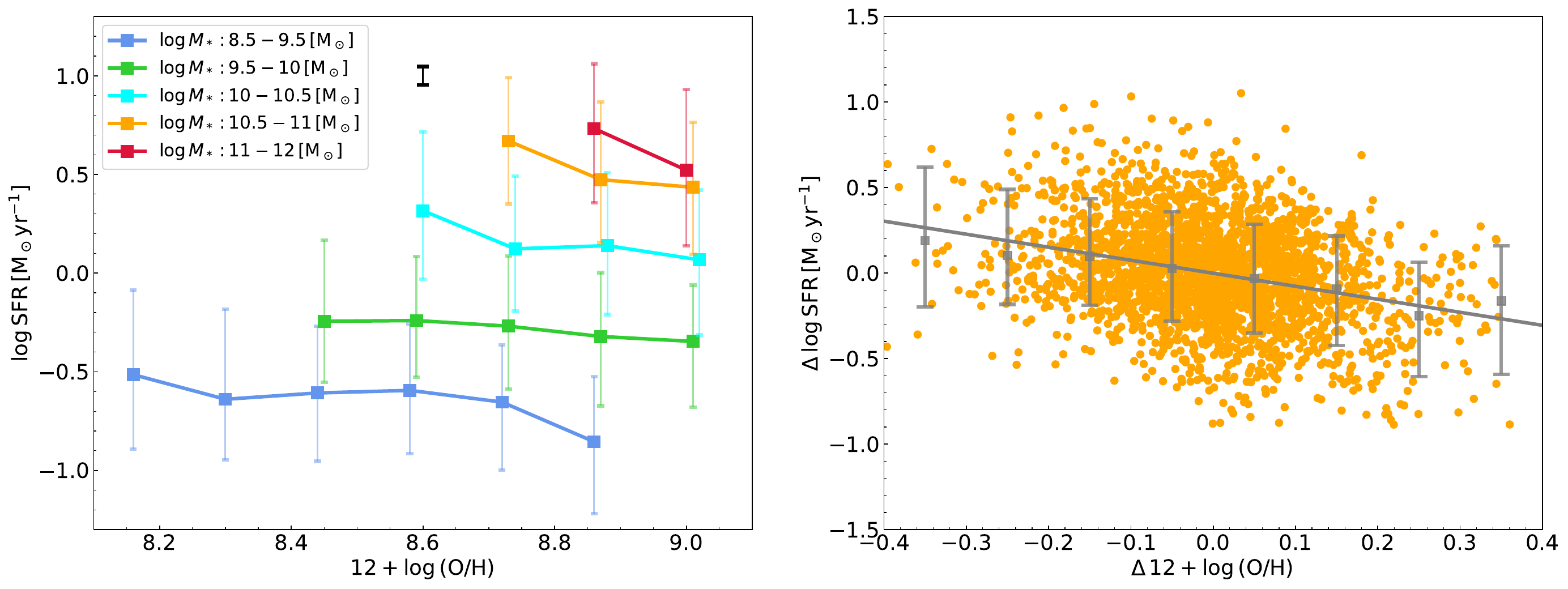}}
    \caption{\textbf{Left:} $\log \mathrm{SFR}$ versus $\Oabu$ for BPT-SF galaxies divided into five stellar mass bins. Points show median values, and the error bars show the 1$\sigma$ scatter of $\log \mathrm{SFR}$ in each oxygen abundance bin; note that the oxygen abundance bins are in fact the same for each stellar mass bin, but the points have been offset slightly for readability. by contrast, the black error bar on the top of the panel shows the typical uncertainty on the median $\log \mathrm{SFR}$ in each bin, which is much smaller than the scatter due to the large number of galaxies per bin. We see that in each stellar mass bin, $\log \mathrm{SFR}$ and $\Oabu$ are anti-correlated. \textbf{Right:} $\Delta\! \log \mathrm{SFR}$ versus $\Delta\, \Oabu$ for the same galaxies as shown in the left panel, where $\Delta\, q$ indicates the difference between some quantity $q$ for a particular galaxy and the mean value for galaxies of matching stellar mass. The grey squares and error bars show medians and 1$\sigma$ scatters in bins of $\Delta\, \Oabu$. The line shows a linear fit to the data, with a best-fit slope of $-0.76$.}
    \label{fig:sfr_dep}
\end{figure*}

\begin{figure}
{\includegraphics[width=\columnwidth]{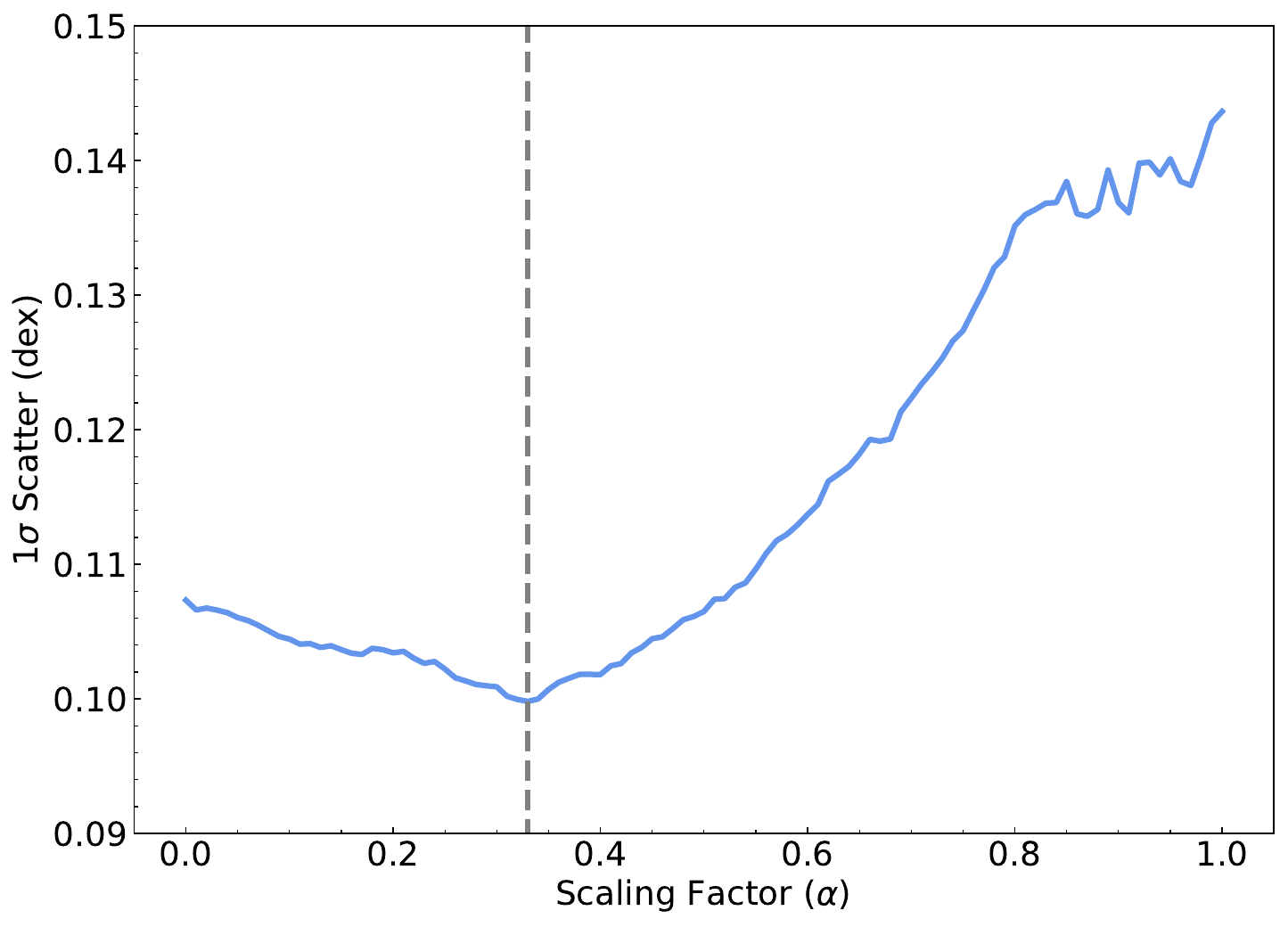}}
    \caption{The relation between the oxygen abundance residual scatter after removing the dependence on $\mu$ in $\mu=\log M_*-\alpha \log \mathrm{SFR}$ and the scaling factor $\alpha$ in BPT-SF galaxies. The dashed vertical line indicates $\alpha$ of the minimum scatter, which is 0.33.}
    \label{fig:scaling}
\end{figure}

The color coding in \autoref{fig:mzr_sf} shows $\log \mathrm{SFR}$ for each galaxy. The most obvious trend visible in the colours is that SFR increases with stellar mass, as a result of SFMS. However, there is an important secondary trend visible, which is that isocolour regions are tilted relative to vertical, indicating a secondary correlation between SFR and oxygen abundance at fixed stellar mass -- i.e., that the BPT-SF galaxies in our sample obey an FZR in addition to an MZR. To explore this dependence more directly, in the left panel of \autoref{fig:sfr_dep} we show $\log \mathrm{SFR}$ versus $\Oabu$ for our BPT-SF sample divided into five stellar mass bins from $10^{8.5} - 10^{12}$ M$_\odot$; the central three bins are 0.5 dex wide, and the smallest and largest are 1 dex wide to compensate for the smaller number of data points at the lowest and highest masses. We estimate the scatter in this relation with an approach analogous to the one we use in \autoref{fig:mzr_sf}: we bin the data by $\Oabu$, and in each bin we compute the 1$\sigma$ scatter of $\log \mathrm{SFR}$ by bootstrap sampling from the Gaussian uncertainty distributions for $\log \mathrm{SFR}$ for each galaxy that falls into that bin by 1,000 times. The uncertainty on an individual-galaxy $\log \mathrm{SFR}$ measurement is the combination of the measurement uncertainty within 1 $\re$, calculated by propagating the error on the H$\alpha$ flux and $f_\mathrm{HII}$ through \autoref{equ:sfr_re}, and the intrinsic scatter we introduce when converting $\log \mathrm{SFR}_{\re}$ to $\log \mathrm{SFR}$ of the whole galaxy using our aperture correction, \autoref{equ:sfr_whole}. The latter is always dominant owing to the high signal-to-noise ratio of H$\alpha$ emission lines.

We find that in all five stellar mass bins the SFR is negatively correlated with oxygen abundance, albeit with large scatter ($\sim$ 0.35 dex). We emphasise that this measurement is much stronger than the large error bars in \autoref{fig:sfr_dep} might at first suggest, because the error bars we are showing represent the \textit{scatter} in each bin, \textit{not} the uncertainty on its central value, which is much smaller due to the large number of galaxies per bin. To illustrate this we estimate the uncertainty on the median in each oxygen abundance bin by bootstraping: given an oxygen abundance bin with $n$ galaxies, we randomly select $n$ galaxies from that bin with replacement (i.e., the same galaxy may be drawn multiple times), and for each galaxy drawn we then draw a value of $\log \mathrm{SFR}$ from the uncertainty distribution on $\log\mathrm{SFR}$ for that galaxy. The median value of these $n$ values is one realisation. We iterate this procedure 1,000 times to compute the scatter of the median of 1,000 realisations (rather than the full scatter of the resulting sample, which is what the coloured error bars show). We show the median uncertainty over all oxygen abundance bins as the black error bar in the top left panel of \autoref{fig:sfr_dep}. Compared to this error bar, the correlation we measure between $\Oabu$ and $\log \mathrm{SFR}$ at fixed stellar mass is clearly highly significant.

To quantify the significance of the correlation, we perform a $t$-test comparing the SFR distributions of the lowest and highest oxygen abundance bins for each stellar mass bin. Specifically, the null hypothesis for this test is that the mean $\log \mathrm{SFR}$ in the lowest $\Oabu$ bin is \textit{smaller} than or equal to that in the highest $\Oabu$ bin, rather than larger as the plot appears to suggest. The p-values returned by this test, from the lowest to highest stellar mass bins, are $5.7\times 10^{-5}$, $\rm 8.7\times 10^{-3}$, $4.1\times 10^{-4}$, $1.2\times 10^{-3}$, and $5.1\times 10^{-4}$, meaning that we can rule out the null hypothesis, and we therefore detect an anti-correlation between $\Oabu$ and $\log \mathrm{SFR}$ at fixed stellar mass, with high confidence. The relation between $\log \mathrm{SFR}$ and $\Oabu$ is flattest, and the $p$-value weakest, in the bin of $\log (M_*/\Msun) \in [9.5, 10]$, even though a clear tilt in isocolour region remains visible in this stellar mass range in \autoref{fig:mzr_sf}. We suspect that the steep relation between oxygen abundance and stellar mass within this mass range gives rise to a positive correlation with SFR that partially cancels the anti-correlation between SFR and oxygen abundance at fixed stellar mass, so that using finer stellar mass binning would yield a more significant result. However, even without this step, the p-value for this bin is still $<10^{-2}$, highly significant.

Given the evidence for an FZR in the BPT-SF sample, we next seek to quantify it. As a first step in this direction, in the right panel of \autoref{fig:sfr_dep}, we show $\Delta\! \log\mathrm{SFR}$ versus $\Delta\, \Oabu$, where we define $\Delta\! \log\mathrm{SFR}$ as the difference between the SFR of an individual galaxy and the mean SFR of galaxies of matching stellar mass, and similarly for $\Delta\, \Oabu$. A negative trend is again evident, and a linear fit to the data gives a slope of $-0.76$. This approach suggests one way of defining an FZR. However, we also consider the approach suggested by \cite{mannucci2010}, whereby we define a parameter combining the stellar mass and SFR: $\mu=\log M_*-\alpha \log \mathrm{SFR}$. The minus symbol in front of $\log \mathrm{SFR}$ suggests the negative correlation between SFR and oxygen abundance, and thus we expect the scaling factor $\alpha$ to lie in the range from 0 to 1. To determine the value of $\alpha$, we construct a data set consisting of ($\mu, \Oabu)$ pairs for some chosen value of $\alpha$, and then use the {\tt scipy.interpolate.splrep} routine to construct a third-order basis spline representation of the data with a smoothing parameter $s=50$, where $s$ controls the interplay between the smoothness of the resulting curve and the quality of the approximation of the data; $s=50$ is a high-level of smoothness, and thus the resulting curve is close to a single best-fit polynomial. We then subtract this smoothed relation from the data and measure the 1$\sigma$ scatter.  We plot this scatter as a function of $\alpha$ in \autoref{fig:scaling}, which shows a clear minimum in the scatter for $\alpha=0.33$. This scatter in the FZR computed using $\alpha=0.33$ is smaller than the scatter in the MZR by $\approx 0.01$ dex. While this number is not large, given the large sample size it is significant. To demonstrate this we carry out an $F$-test on the distributions of the residual scatter of MZR and FZR with the $\alpha=0.33$. The null hypothesis is that the standard deviation of residuals from the MZR is smaller than or equal to that of the FZR. The resulting p-value is $2.6\times 10^{-3}$, indicating that the FZR reduces the scatter relative to the MZR with high statistical significance.

\subsection{MZRs for AGN-Host Galaxies}
\label{sec:other_gal}

\begin{figure*}
    \resizebox{17cm}{!}{\includegraphics{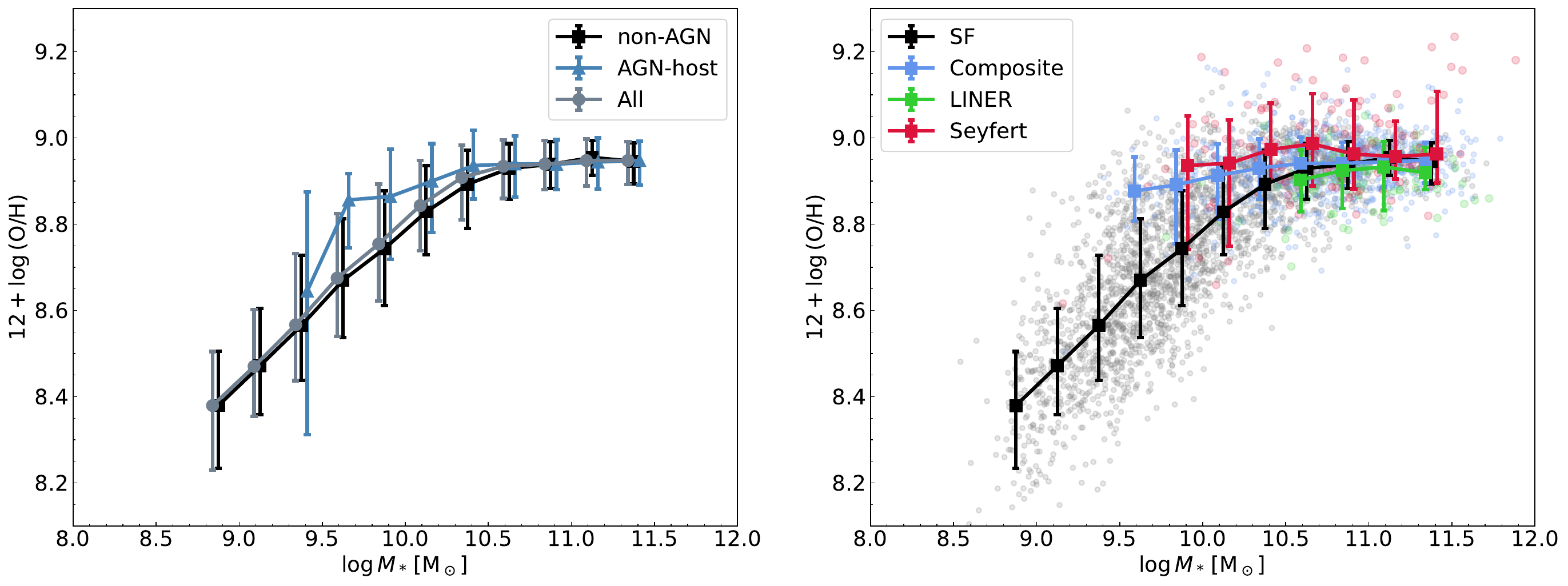}}
    \caption{\textbf{Left:} The median oxygen abundance and 1$\sigma$ level scatter of non-AGN (\textbf{SF}) galaxies (black), AGN-host galaxies (blue), and all galaxies combining non-AGN and AGN hosts (grey); note that the mass bins used are the same for each sample, but the horizontal locations of the points shown have been offset slightly to improve readability. AGN-host galaxies are biased to higher oxygen abundance at lower stellar masses, with a maximum deviation around $10^{9.6} \Msun$, but otherwise follow a similar trend to non-AGN galaxies. \textbf{Right:} The MZR for Composite (blue), LINER (green), and Seyfert galaxies (red), compared to that of BPT-SF (non-AGN) galaxies (black); coloured background points show individual galaxies, while lines and the error bars are the medians and the 1$\sigma$ level scatter in each stellar mass bin. Note that differences in the sizes of points representing individual galaxies are only for visibility, and have no physical meaning. The Composite and Seyfert galaxies have similar oxygen abundance at high mass end but are biased to higher oxygen abundance at low mass bins.}
    \label{fig:mzr_other}
\end{figure*}

Now that we have developed our analysis toolkit for non-AGN (BPT-SF) galaxies, we proceed to apply the same methods to the AGN-host galaxies, starting with the MZR in \autoref{fig:mzr_other}. The left panel of this figure compares all AGN-host galaxies grouped together (including Ambiguous) to both non-AGN (BPT-SF) galaxies alone and to the whole sample including both non-AGN and AGN hosts, while the right panel further breaks out the MZRs for the AGN-host sample by decomposing it into Composite, LINER, and Seyfert galaxies.  Our methods for computing both individual data points and binned means and uncertainties are the same for each sub-class shown, and completely parallel the methods used to construct \autoref{fig:mzr_sf}, and thus all samples are directly comparable.

First focusing on the left panel, we find that non-AGN and AGN-host galaxies have nearly identical MZRs at stellar masses $\gtrsim 10^{10.5}$ $\rm \Msun$, where both sub-classes lie on the flat part of the MZR. The two samples begin to diverge slightly at lower stellar masses, with both the mean and the upper and lower envelopes of the distribution lying at higher oxygen abundance in the AGN-host galaxies. The difference is largest at around $10^{9.6} \Msun$, where AGN hosts have mean oxygen abundance that are $\approx 0.2$ dex larger than non-AGN galaxies of equal mass; it is unclear if the trend continues at even lower stellar masses, since the uncertainties in the AGN-host sample become increasingly large due to the small sample size. At all masses we find that the MZR of all galaxies (including both non-AGN and AGN hosts) is nearly identical to that of non-AGN galaxies alone, which is not surprising given that non-AGN host galaxies our number AGN hosts roughly $2.5:1$ in our sample.

Turning to the right panel of \autoref{fig:mzr_other}, we see that the general pattern that non-AGN and AGN hosts are nearly identical at high stellar mass continues to hold for each of the AGN-host sub-types. LINER, Seyferts, and Composite galaxies differ mainly in the stellar mass at which they begin to be biased to higher oxygen abundance than BPT-SF galaxies, and in the extent of the bias once it appears. Seyfert galaxies lie farthest above the MZR defined by BPT-SF galaxies, and begin to deviate noticeably from it even at stellar masses approaching $10^{11}$ M$_\odot$, well into the flat part of the MZR. By contrast, LINER appear to have slightly smaller oxygen abundance than BPT-SF galaxies out to the lowest stellar masses for which there are appreciable samples (which admittedly is only down to $\approx 10^{10.5}$ M$_\odot$). Composite galaxies lie in between these extremes, showing less deviation than Seyferts and more than LINERs, and beginning to deviate from the BPT-SF galaxy MZR at lower masses than Seyferts as well.

\subsection{FZRs for AGN-Host Galaxies}
\label{sec:fzr}

\begin{figure}
    {\includegraphics[width=\columnwidth]{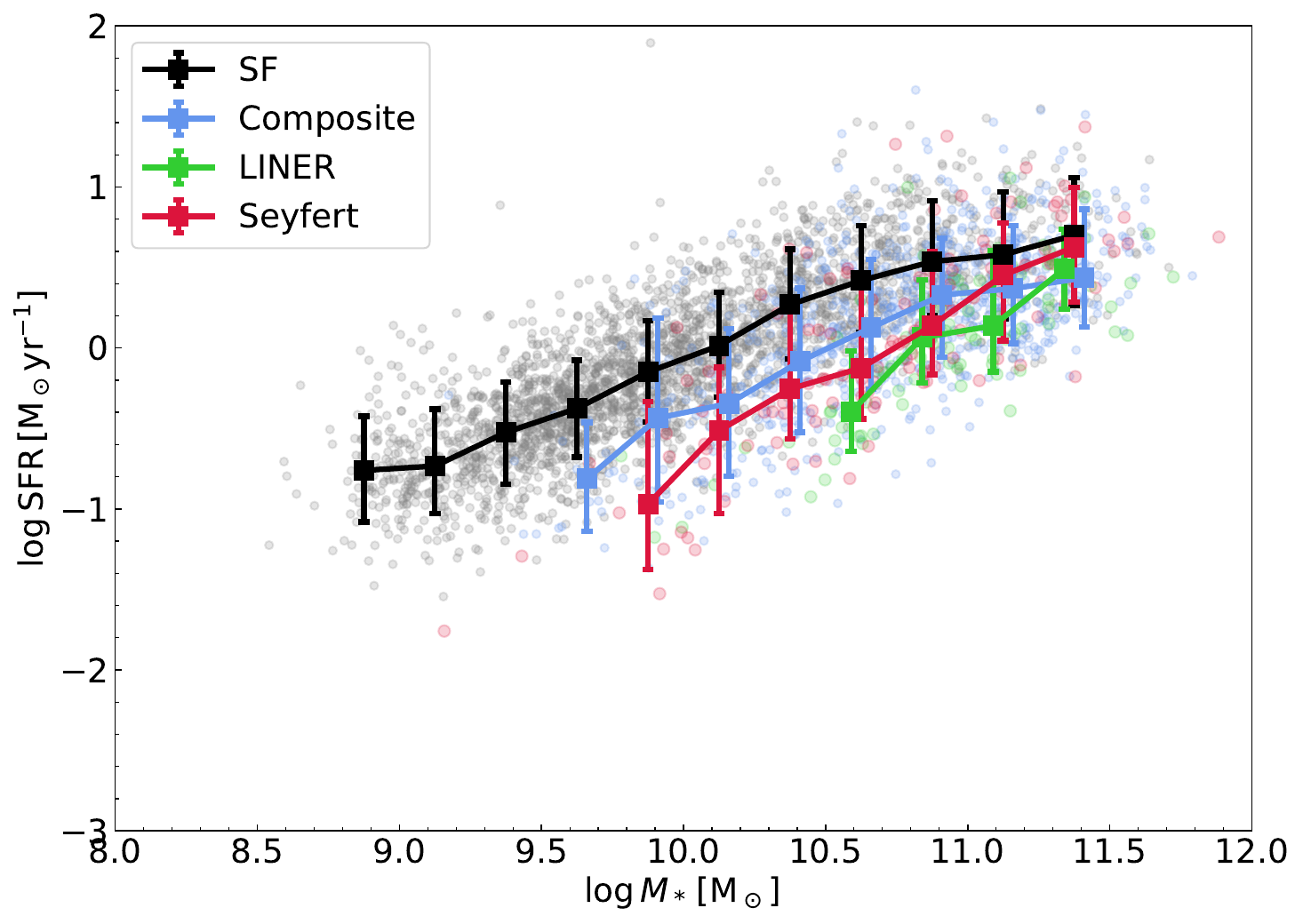}}
    \caption{The $\log \mathrm{SFR}$ versus $\log M_*$ relation for BPT-SF (black), Composite (blue), LINER (green), and Seyfert galaxies (red). The lines and the error bars are the medians and the 1$\sigma$ level scatter in each stellar mass bin. AGN-host galaxies generally have lower SFR than BPT-SF galaxies and thus are off SFMS.}
    \label{fig:sfms}
\end{figure}

\begin{figure*}
    \resizebox{17cm}{!}{\includegraphics{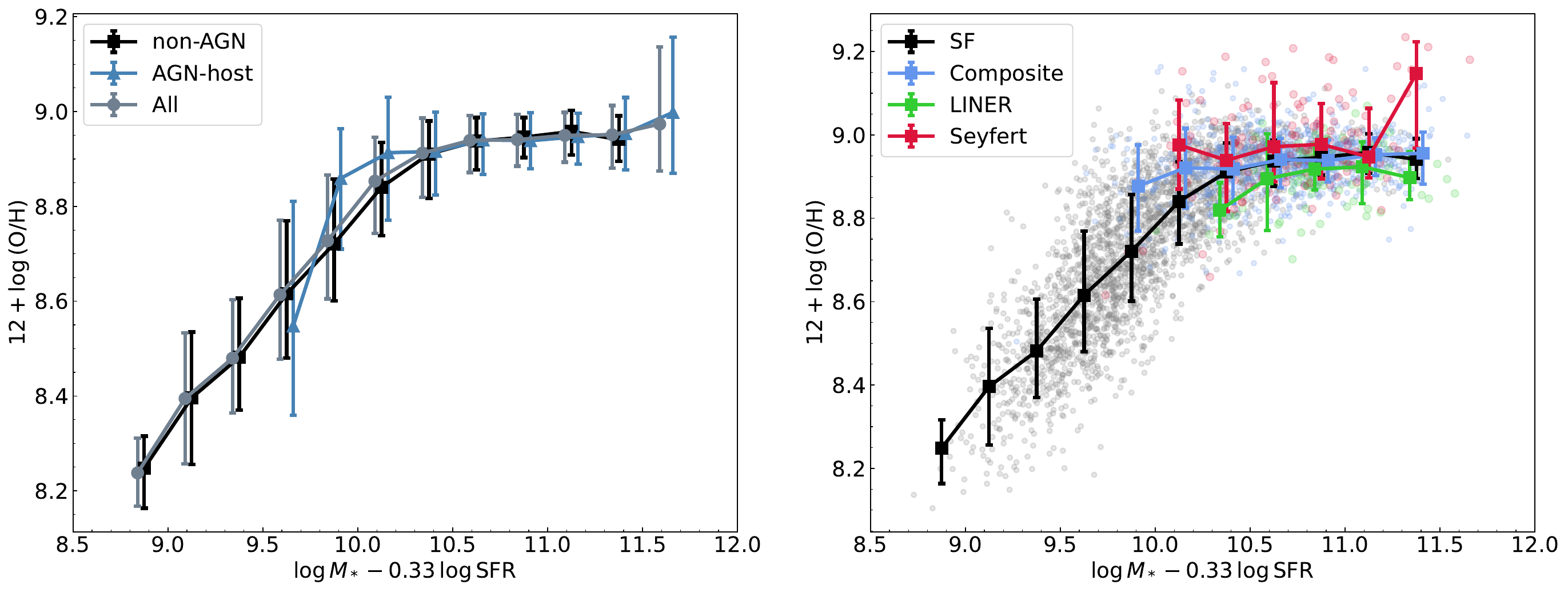}}
    \caption{The same as \autoref{fig:mzr_other}, but changing the horizontal axis from $\log M_*$ to $\mu=\log M_*-\alpha \log \mathrm{SFR}$ with $\alpha=0.33$.}
    \label{fig:fzr}
\end{figure*}

To investigate the FZR for AGN-host galaxies, we firstly explore the relation between SFR and stellar mass. NebulaBayes allows us to disentangle the relative contributions to the H$\alpha$ emission line from HII regions and NLRs, which in turn enables us to estimate the SFR for AGN-host galaxies by using $f_\mathrm{HII}$ corrected H$\alpha$ luminosity. This decomposition will become interesting for the purposes of understanding the FZR for AGN-host galaxies. \autoref{fig:sfms} shows $\log \mathrm{SFR}$ versus $\log M_*$ for Composite, LINER, and Seyfert galaxies compared with the SFMS of the BPT-SF sample. We find that AGN-host galaxies generally have reduced SFR and are moving off the SFMS into the green valley. We also find a difference between the slopes of the different AGN-host subtypes in the $\log\mathrm{SFR}-\log M_*$ plane: the Composite galaxies generally have a slope similar to that of BPT-SF galaxies, but Seyfert and LINER galaxies have steeper slopes so that they lie closer to the SFMS at high masses and deviate more strongly from it at lower masses.

Keeping this trend in mind, in \autoref{fig:fzr} we plot the FZR of AGN-host and non-AGN galaxies, using the same parameter $\mu = \log M_* - 0.33 \log\mathrm{SFR}$ that we measured for the BPT-SF sample alone. As with \autoref{fig:mzr_other}, the left panel shows all AGN-host galaxies (including Ambiguous) and the whole sample combining both non-AGN and AGN hosts, while the right panel breaks up the AGN-host group and shows the FZR for Composite, LINER, and Seyfert galaxies along with BPT-SF galaxies. A clear trend that is visible in both panels is that the offset between AGN-host and non-AGN galaxies in the FZR (\autoref{fig:fzr}) is smaller than in the MZR (\autoref{fig:mzr_other}).

It is straightforward to understand why this occurs: AGN-host galaxies generally have reduced SFR comparing with BPT-SF galaxies, so AGN-host galaxies with the same stellar mass as BPT-SF have larger $\mu$ and thus move toward the right in the FZR compared to the MZR. Due to the flatness of MZR at the high mass end, and the similarity of MZR between non-AGN and AGN-host galaxies, the FZR of AGN-host galaxies still overlaps with that of non-AGN galaxies at high stellar mass, i.e., at high stellar mass, where AGN and non-AGN galaxies already have similar MZRs, this shift in going from the MZR to the SFR does not create a difference between AGN and non-AGN galaxies because the shift is parallel to the existing relation. In \autoref{fig:fzr}, this description applies at roughly $\mu \gtrsim 10.4$. By contrast, at lower stellar masses where the MZR is not flat, AGN-host galaxies lie above and to the left of non-AGN galaxies in the MZR -- at fixed stellar mass, they have higher oxygen abundance. However, the AGN hosts also have lower SFR, so when we go from the MZR to the FZR and the AGN-host galaxies shift right, this brings them closer to the non-AGN galaxies. Quantitatively, we say that the largest difference between AGN- and non-AGN galaxies in the MZR was at a stellar mass $\log (M_*/\Msun) = 9.6$, where AGN-host galaxies have a $\approx 0.2$ dex higher oxygen abundance. When we go to the FZR, the maximum offset between AGN hosts and non-AGN galaxies occurs at $\mu = 9.9$, and is reduced to $\approx 0.13$ dex. The same trend also happens for each AGN sub-type: the maximum offsets between BPT-SF and Composite, Seyfert galaxies seen in the MZR are reduced to $\approx 0.16$ and $\approx 0.13$ dex, respectively. The LINER galaxies still lie slightly below the BPT-SF galaxies as in MZR.

While this trend strongly hints that AGN-host galaxies lie on the same FZR as non-AGN galaxies, interestingly we do not find evidence for an FZR in the AGN-host sample as we do for the non-AGN. Quantitatively, we can compare the scatters in the MZR and the FZR for the AGN-host sample using the same $F$-test we used for the BPT-SF galaxies. While this exercise for the BPT-SF galaxies indicated that the FZR has less scatter than the MZR with high statistical significance, the same is not true of the AGN-host sample: the p-values we obtain are 0.49, 0.54, and 0.41 for the Composite, LINER, and Seyfert galaxies, respectively, and doing the same test for all of the AGN-host galaxies combined together yields a p-value of 0.79. This indicates that the reduction in scatter in going from the MZR to the FZR is not statistically significant for AGN-host galaxies. We defer a discussion of how to reconcile these seemingly contradictory results -- that AGN-host and non-AGN galaxies lie closer to one another on the FZR than on the MZR, but that there is no evidence for an FZR in the AGN-host sample -- to \autoref{sec:dis_fzr}.

\section{Discussion}
\label{sec:discussion}

\begin{figure*}
\resizebox{17cm}{!}{\includegraphics{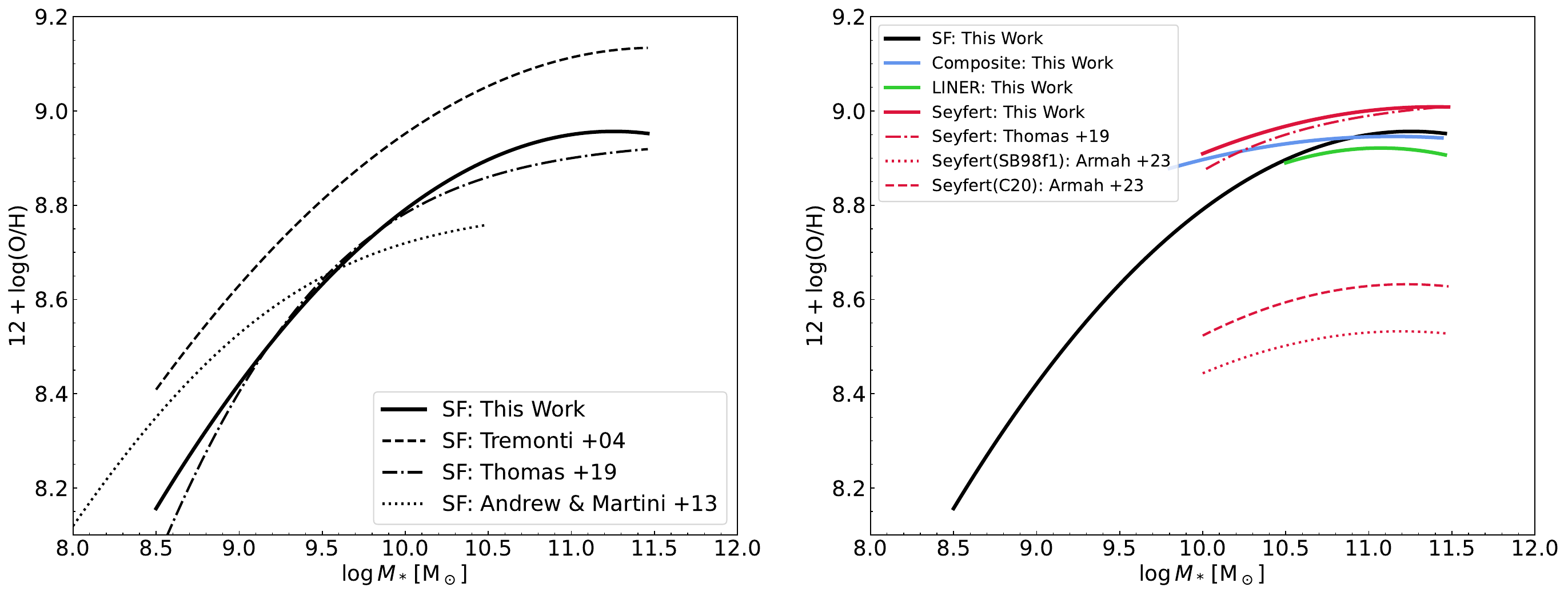}}
    \caption{\textbf{Left:} Comparison between the MZR for BPT-SF galaxies we derive in this work (black solid line; \autoref{equ:mzr}) with literature results from \citet[dashed line, their Equation 3]{tremonti2004}, \citet[dot-dashed line, their Equation 1 using coefficients from their Table 1]{thomas2019}, and \citet[dotted line, their Equation 5 using coeffficients from their Table 4]{andrews2013}. \textbf{Right:} Same as left panel, but for AGN-host galaxies. We show the second-order polynomial fits to the MZRs derived in this work for Composite (blue), LINER (green), and Seyfert galaxies (red). The red dashed line is the relation of Seyfert galaxies reported in \citet{thomas2019}, and the red dashed and dotted lines are the MZRs for Seyfert galaxies from \citet{armah2023} using the \citet[][C20]{carvalho2020} and \citet[][SB98f1]{storchi1998} calibrators, respectively. The black solid line is the MZR of BPT-SF galaxies derived in this work for reference, and is the same as the black solid line in the left panel.}
    \label{fig:mzr_compare}
\end{figure*}

In this section, we first compare our results of MZR to those in the literature in \autoref{sec:dis_mzr_sf}, and then in \autoref{sec:dis_fzr} we discuss possible interpretations of our results for the FZR in AGN-host galaxies. We finally discuss possible future improvements to our analysis in \autoref{sec:dis_ingred}.

\subsection{Comparing MZR to Literature Results}
\label{sec:dis_mzr_sf}

\subsubsection{Non-AGN galaxies}

We firstly compare the MZR of non-AGN (BPT-SF) galaxies in this work with the literature. The left panel of \autoref{fig:mzr_compare} shows our second-order polynomial fit to the MZR (\autoref{equ:mzr}; black solid line) together with the fits obtained by \citet[dashed line]{tremonti2004}, \citet[dash-dot line; also derived using NebulaBayes, but on spatially-unresolved data]{thomas2019}, and \citet[derived using a direct $T_e$-based method]{andrews2013}. Our MZR is most similar to that of \cite{thomas2019}, which is not surprising since both are derived from NebulaBayes. The differences that exist are partly due to the different functional forms used to fit the data (\autoref{equ:mzr} versus Equation 1 in their work), but also arise due to aperture effects and because our grid of HII region models differs from theirs in several aspects (e.g., geometry, stellar tracks, and duration of star formation). Compared to \cite{tremonti2004}, we recover a similar low-mass slope and turnover, but our MZR is offset to lower absolute abundances by $\approx 0.15$ dex. \citet{thomas2019} propose that this discrepancy is due to modeling: \citeauthor{tremonti2004}'s models are calibrated to observed line ratios from nearby SF galaxies, while models used in NebulaBayes are purely theoretical. However, the models used in \citeauthor{tremonti2004} are calibrated in \citet{charlot2001}, who plot several line ratios from models against local spiral, irregular, starburst galaxies, and HII regions to select the standard values and upper and lower limits of parameters (e.g., effective metallicity, ionization parameter, dust attenuation, and dust-to-metal ratio). When estimating the oxygen abundance of individual galaxies, we use similar procedures, i.e.: Bayesian methods from pure theoretical models. Thus the underlying reason for the discrepancy is most likely to be improvements to atomic data, metal scaling relations, and stellar synthesis codes that have occurred over the intervening two decades.

The comparison of work with the results of \cite{andrews2013} is particularly interesting, because they derive their oxygen abundance using a $T_e$-based approach, which is generally believed to be the most reliable way to estimate the oxygen abundance \citep{kewley2019a}. In general $T_e$ methods give systematically lower abundances than strong-line methods \citep{maiolino2019}. The $\mathrm{[NII]/[OII]}$ prior that we adopt in this work is essentially a strong-line method. It is therefore somewhat surprising that our MZR has a similar normalization to the one obtained by \citeauthor{andrews2013}, although there are differences in detail: the \citeauthor{andrews2013} sample yields higher oxygen abundance at lower stellar mass and the opposite at moderate stellar mass\footnote{\citet{andrews2013}'s results do not extend above $\approx 10^{10.5}$ M$_\odot$ due to the difficulty in applying $T_e$-based methods to galaxies with high oxygen abundance.}, thus leading to a lower overall slope. The origin of this effect is difficult to address because the ultimate cause of the offset between $T_e$-based and strong-line oxygen abundance calibrations remains debated \citep{kewley2008}, though the simplified geometry generally adopted in HII regions models remains a likely culprit \citep[e.g.,][]{jin2022a,jin2022b}. 

Our results for the FZR of non-AGN galaxies is also broadly compatible with previous work. We find that an FZR using $\alpha = 0.33$ leads to the smallest residual scatter, similar to the value of 0.32 found by \cite{mannucci2010}. However, the amount by which we reduce the scatter compared to the MZR, while statistically significant, is only 0.01 dex, while \citeauthor{mannucci2010} find a larger reduction of 0.035 dex. Our result is closer to that of \cite{perez2013}, who also find a 0.01 dex reduction in the scatter when going from the MZR to the FZR in SDSS-DR7. As suggested by \cite{salim2014}, the larger scatter reduction found by \cite{mannucci2010} is likely due to their use of binned rather than raw data, as was the practice in later work, and our results are compatible with that conclusion.

\subsubsection{AGN-host galaxies}

The MZR for AGN-host galaxies has received far less attention in the literature, and the FZR -- which we defer discussing to \autoref{sec:dis_fzr} -- none at all. We can, however, compare to what results there are. The red dashed and dotted lines in the right panel of \autoref{fig:mzr_compare} show the MZRs of local Seyfert galaxies derived by \cite{armah2023} using the calibrators proposed by \citet[][C20]{carvalho2020} and \citet[][SB98f1]{storchi1998}, respectively. Our MZR for Seyferts has larger oxygen abundance by 0.37 dex and 0.46 dex. Although some of this offset is likely due to the different metallicity calibrators used, since as for SF galaxies the discrepancies between different calibrators are significant (as evidenced by the 0.09 dex offset between the C20 and SB98f1 calibrations within the \citeauthor{armah2023} sample), a more important contributor is the assumed abundance scaling of different elements. We can understand this effect taking C20 as an example. In this method, the Solar oxygen abundance is taken to be 8.69, and the abundances of other heavy elements are linearly scaled with the total metallicity $Z$. In contrast, as shown in \aref{sec:mappings}, we use Galactic Concordance abundances from \cite{nicholls2017}, which take the ``Solar'' oxygen abundance to be 8.76 -- the average oxygen abundance of 29 local B stars -- and then uses an observationally-motivated and more realistic model for how the relative abundances of other elements vary with total metallicity $Z$; this different abundance scaling modifies the predicted relationship between metallicity and HII region line ratios. A second and related contribution to the offset is the assumed nitrogen-to-oxygen ratio, which is important in both this work and C20. C20 take their ratio from \cite{pilyugin2016}, who find an oxygen abundance at Solar N/O that is 0.33 dex lower than our value from \cite{nicholls2017}.

By contrast our MZR for Seyfert galaxies is generally in good agreement with the previous result from \citet{thomas2019}, which is not surprising since we are using the same tool (NebulaBayes) and thus ultimately the same set of assumed abundance scalings and line diagnostics. It is clear from the \autoref{fig:mzr_compare} that the systematic uncertainties associated with these choices are much larger than those associated with other differences between our data and procedure and those used by \citeauthor{thomas2019}, including size of aperture, spatially-resolved versus unresolved data, and method of fitting a functional form for the MZR to the data.

\subsection{On the Fundamental Metallicity Relation in AGN-Host Galaxies}
\label{sec:dis_fzr}

\begin{figure}
{\includegraphics[width=\columnwidth]{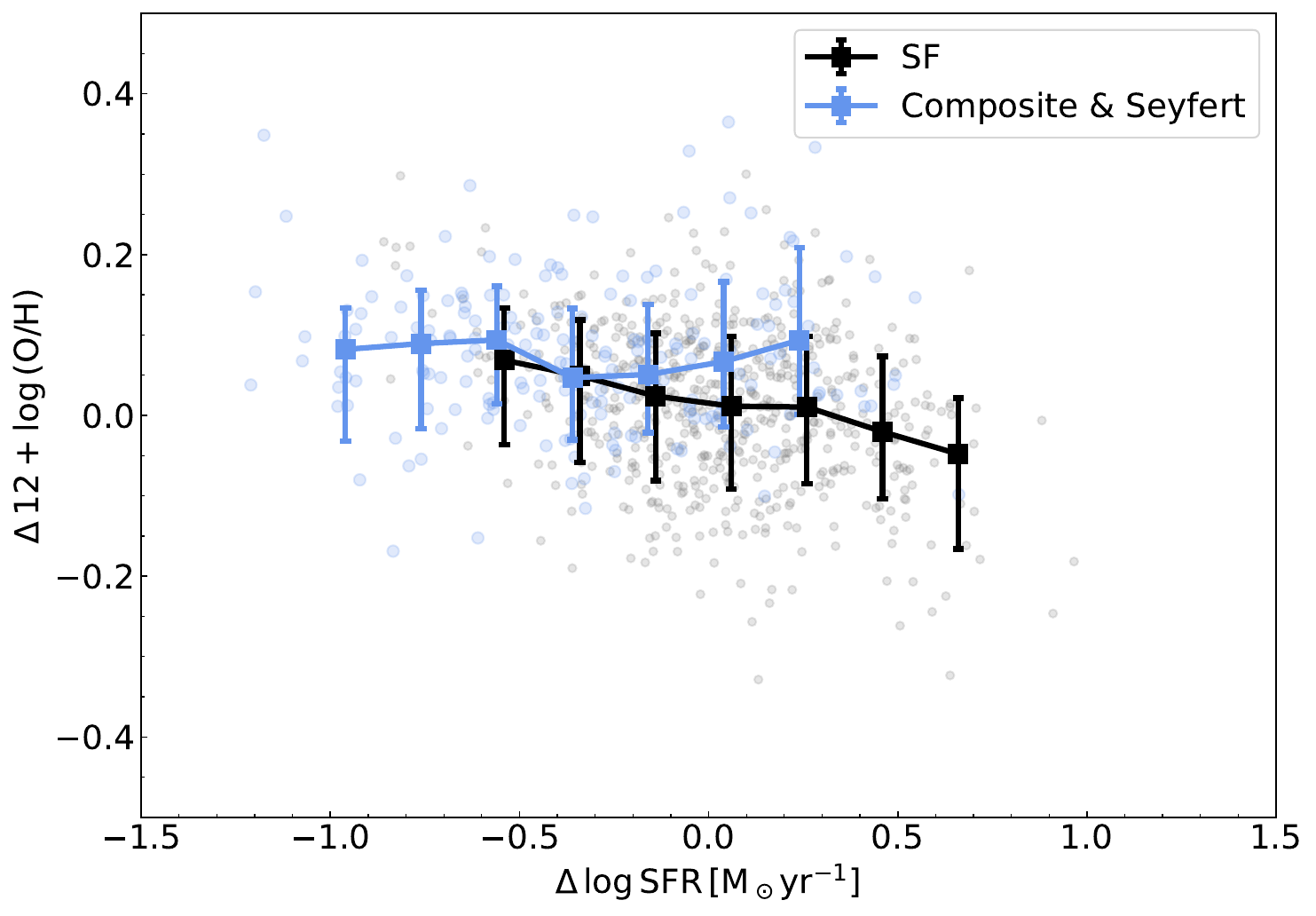}}
    \caption{$\Delta\, \Oabu$ versus $\Delta\! \log \mathrm{SFR}$ relation for BPT-SF galaxies (black), and a combination of Composite and Seyfert galaxies (blue) within $10^{10} - 10^{10.5} \Msun$, where $\Delta\, q$ indicates the difference between some quantity $q$ for a particular galaxy and the mean value for BPT-SF galaxies of matching stellar mass. The median values and $1\sigma$ scatter levels in each $\Delta\! \log \mathrm{SFR}$ bin are also presented. Non-SF galaxies generally have lower $\Delta\! \log \mathrm{SFR}$ and do not extend the negative trend of BPT-SF galaxies.}
    \label{fig:dabu_compare}
\end{figure}

Our work is the first published exploration of the FZR in AGN-host galaxies, and yields the interesting result that, while AGN hosts are displaced substantially from non-AGN galaxies on the MZR around $10^{9.5} \Msun$ (\autoref{fig:mzr_other}), they are much less displaced on the FZR (\autoref{fig:fzr}), and this effect is much larger than the typical reduction in scatter in going from the MZR to the FZR in non-AGN -- we find that for non-AGN the scatter in the FZR is smaller than that in the MZR by only $\approx 0.01$ dex, while in the FZR the offset between AGN- and non-AGN-host galaxies is reduced by as much as $\approx 0.07$ dex compared to the MZR. In a purely mechanical sense we can attribute the offset between non-AGN hosts and AGN hosts on MZR to the fact that AGN hosts have systematically lower SFRs than non-AGN galaxies of equal stellar mass. Indeed, this effect remains even though there is a bias toward higher SFR due to our selection criteria: most of the AGN-host galaxies that could potentially be in our sample are excluded by the minimum H$\alpha$ equivalent width that we require, an effect that is most prominent in LINER galaxies that contain large areas that are likely ionised by pAGB stars, yielding typical H$\alpha$ EWs $\sim 1-2$ \AA~\citep{belfiore2016}. Even with this bias, our AGN hosts are still show reduced SFR, and this is consistent with the findings in previous works that the optically selected AGN generally have low SFR comparing with galaxies on SFMS \citep{ellison2016}, and with the proposal that AGN feedback suppresses the SF activities \citep[e.g.,][]{hopkins2008,hopkins2010,bing2019}. 

However, this does not yet answer the question of why, despite this result, we do not find any compelling statistical evidence for an FZR for AGN-host galaxies, and more broadly, how to interpret the relationship between the MZR and FZR for AGN-host galaxies versus all other BPT-SF galaxies. To help make sense of this, in \autoref{fig:dabu_compare} we show $\Delta\, \Oabu$ versus $\Delta\! \log \mathrm{SFR}$ for both BPT-SF and AGN-host galaxies, where for the AGN-host galaxies we limit the sample to Seyfert and Composite galaxies, with stellar masses in the range $10^{10} - 10^{10.5} \Msun$, which are the galaxies that are the most displaced from BPT-SF galaxies in the MZR (c.f.~\autoref{fig:mzr_other}). The reference zero points of both $\Oabu$ and SFR in each stellar mass bin are the medians of BPT-SF galaxies.

This figure makes it clear how to reconcile our seemingly contradictory results. First, we find that the median location of AGN-host galaxies in $\Delta\, \Oabu$ versus $\Delta\! \log \mathrm{SFR}$ coincides well with the low $\Delta\! \log \mathrm{SFR}$ end of the non-AGN population, and has a oxygen abundance that is higher than the mean by a similar amount. This is consistent with our proposal in \autoref{sec:fzr} that the offset between non-AGN and AGN-host galaxies is a natural result of the existence of the FZR: AGN-host galaxies generally have reduced SFR compared to non-AGN galaxies, and this reduction in SFR seems to lead to an increase in oxygen abundance by about the same amount as in non-AGN galaxies. 

However, at the same time it is clear from \autoref{fig:dabu_compare} that the $\Delta\, \Oabu$ versus $\Delta\! \log \mathrm{SFR}$ relation of AGN-host galaxies is quite flat, rather than showing the same negative trend as BPT-SF galaxies. This is why the F-tests on AGN-host galaxies show no significant scatter reduction from MZR to FZR. Although the AGN-host population as a \textit{whole} is displaced to both lower SFR and higher oxygen abundance than the mean of non-AGN galaxies, there is no correlation between SFR and $\Oabu$ \textit{within} AGN-host galaxies, and thus no FZR internal to that population. One possible explanation for this phenomenon is the proposal by \cite{salim2014} that the dependence of oxygen abundance on SFR is strongest in extreme SF galaxies and is explained by the scenario proposed by \citeauthor{mannucci2010} that the SFR-oxygen abundance correlation at fixed stellar mass arises from events where galaxies accrete substantial amounts of pristine gas, simultaneously diluting the gas phase metallicity and enhancing the SFR. The lack of a high-$\Delta\! \log\mathrm{SFR}$ tail to the AGN-host distribution perhaps suggests that for some reason, perhaps related to AGN feedback, such substantial accretion events do not occur while galaxies host an AGN, known as `starvation' \citep{kumari2021}. In the absence of such major accretion events, the SFR declines for an individual system while its metallicity remains mostly unchanged, yielding the flat rather than sloped relationship between SFR and oxygen abundance for AGN visible in \autoref{fig:dabu_compare}.

\subsection{On the Importance of Uncertainties in the Nitrogen-to-Oxygen Ratio}
\label{sec:caveats}

The oxygen abundances we use in this work, for both BPT-SF and AGN spaxels, are sensitive to the [NII]/[OII] ratio, which enters the NebulaBayes oxygen abundance estimate as a prior (\autoref{sec:NB}). As a result the estimated oxygen abundance is sensitive to the N/O ratio. We assume that the N/O varies with oxygen abundance as found in the Galactic Concordance (GC) abundance set \citep{nicholls2017}, and expressed by \autoref{equ:no_ratio}. In this section, we explore the uncertainties imposed by our adoption of this relation, and the possible effects if the N/O ratio were to vary from our assumed value.

Firstly, let us consider possible effects of gas inflows. Above we concluded that our results support a scenario in which inflow of pristine gas simultaneously reduces the oxygen abundance and enhances the SFR, explaining why oxygen abundance depend on star formation rate at fixed stellar mass. However, as shown in \citet{andrews2013}, such inflow events reduce O/H while leaving N/O largely unaffected, since both elements are diluted by roughly equal amounts. This means that our assumed N/O versus O/H relation, which is calibrated from Milky Way observations, might not apply to galaxies that have recently experienced a strong inflow. However, with a bit of consideration we can see that the error we make by using the GC abundances, if there is one, would actually be causing us to \textit{underestimate} the SFR-dependence of the MZR rather than to overestimate it. To see why, note that if this effect is significant, it means that high sSFR galaxies should in reality have lower O/H ratios at fixed N/O than is expected from GC abundances; thus by adopting the GC abundances when we map the observed [NII]/[OII] ratio to an O/H value, we overestimate the oxygen abundance in these metal-poor, high sSFR galaxies. If we were somehow to correct this bias, the oxygen abundance of our high sSFR galaxies would therefore fall, which would \textit{strengthen} the anti-correlation between the oxygen abundance and SFR at fixed stellar mass that we have found.

Secondly, in this work we adopt the same GC N/O versus O/H relation for both AGN-host and non-AGN galaxies. It is therefore important to ask whether these two galaxy types might have systematically different N/O versus O/H relations, and, if they do, how that would affect our results on the differences between AGN host and non-host galaxies' MZR and FZR. We begin by noting that our approach of assuming the same N/O versus O/H relation for AGN-host galaxies as for non-host galaxies is consistent with most published work on AGN, which for the most part directly adopts N/O versus O/H relations from HII regions or stars \citep[e.g.,][]{carvalho2020,groves2004,ji2020}. There have been only limited efforts to check this assumption: \citet{dors2017} investigate the nitrogen abundance in Seyfert 2 galaxies and find that they obey an N/O versus O/H relation consistent with those of extragalactic HII regions, but we note that this result must be treated with caution because the authors use different abundance calibrators taken for estimating the nitrogen abundance in HII regions and AGNs. That said, there is certainly no direct observational evidence showing a discrepancy in the N/O versus O/H relation between galaxies and do and do not host AGN, and such evidence might be hard to come by given that N/O versus O/H in HII regions has a large scatter (e.g., Figure 3 in \citealt{perez2009}).

In the absence of observational evidence, we can turn to theory to note that variations in N/O at fixed O/H are expected to be driven primarily by galaxy-to-galaxy variations in the star formation history (SFH), resulting in different amounts of secondary versus primary nitrogen, rather than the instantaneous SFR \citep{belfiore2017,nicholls2017,scheafer2020,zhu2023}. To the extent that this expectation is correct, we do not anticipate AGN-host galaxies to have systematically different N/O versus O/H relations than non-hosts, because we do not expect AGN hosts to have SFHs very different from those of non-hosts. AGN reflect a discrete phase with a typical duty cycle $\sim 10^{-4}$, at least in the local Universe \citep{shankar2009}, and all massive galaxies likely go through short AGN phases. This process is stochastic, so that, while few central black holes are active at any given time, those that are active are drawn from the same general galaxy population as those that are quiescent. While this is our theoretical expectation, it is nonetheless clear from this discussion that the large scatter of N/O versus O/H  introduces non-negligible uncertainties in the derived MZR. A more comprehensive relation that models variations in galaxy SFH should be considered in models.

\subsection{Future Improvements}
\label{sec:dis_ingred}

While our analysis method has allowed the determination of the FZR in AGN-host galaxies, and a significant improvement in our knowledge of the MZR for this sample, there remain opportunities for improvement. One would be to add shocks as a potential source of ionised gas emission in our model. Slow shocks can produce LINER-like spectra \citep{rich2011}, while fast shocks ($v > 500 \,\mathrm{km\, s^{-1}}$) drive a photoionizing precursor that can produce Seyfert-like spectra \citep{allen2008}. Gas velocity dispersion greater than $80 \,\mathrm{km\, s^{-1}}$ have been shown to correlate with $\mathrm{[NII]/H}\alpha$, $\mathrm{[SII]/H}\alpha$, and $\mathrm{[OI]/H}\alpha$ ratios, an effect that has already been used in addition to the normal BPT diagrams to disentangle shock, AGN, and SF contributions to emission \citep{dagostino2019a,dagostino2019b,kewley2019a}. However, the low spectral resolution of MaNGA \citep[$\sim$ $70 \,\mathrm{km\, s^{-1}}$,][]{law2021} makes it inefficient for identifying shocks. Higher spectral-resolution surveys such as SAMI \citep[$\sim$ $30 \,\mathrm{km\, s^{-1}}$ in red,][]{scott2018} are a promising avenue for this approach in the future.

Another ingredient we could add to our models is ionisation by pAGB stars. As discussed in \autoref{sec:sample}, we exclude spaxels with H$\alpha$ EW $\leq 3$ \AA~in order to ensure that we apply our models only to regions where ionisation is dominated by massive stars or by AGN, the two ionisation sources modelled in our grids. However, this restriction does noticeably reduce our sample size. By adding pAGB models, we could remove the limitation on the EW of H$\alpha$ and thereby increase the sample size, especially for the LINER group, a significant fraction of which have central regions of low H$\alpha$ EW and are therefore excluded from this study by the requirement that no more than 30\% of the spaxels within $1\re$ of galactic centre be masked. Although we argue that pAGB contamination is not a significant concern given our choice of EW cut, a further minor benefit of including pAGB models in our NebulaBayes grid would be to further mitigate any effects of unmodeled pAGB ionisation.

A third promising direction for future work enabled by our method is investigating the spatially-resolved MZR and FZR. Thanks to the prevalence of IFU surveys, a local, spaxel-scale relationship between oxygen abundance, stellar mass density, and SFR density has been found in SF galaxies \citep{teklu2020,baker2023}, shedding light on the impacts of metal circulation from localized processes in galaxies. Extending this spatially-resolved FZR to AGN-host galaxies would provide us with deeper insights on the interplay between AGN activity and physical processes regulating the FZR. Comparison between the global relations we have studied here and local relations would also help us determine whether the similarity between both the MZRs and FZRs of non-AGN and AGN-host galaxies we have discovered in this work exist due to a similarity between NLRs and HII regions, or whether it is restricted to the HII regions that may still be dominant in AGN-host galaxies. Intriguingly, \citet{donascimento2022} find that the oxygen abundances of NLRs in Seyfert galaxies is lower than HII regions by 0.16-0.3 dex. However, this result relies on using different oxygen abundance indicators for HII regions and NLRs, and it would be useful to verify this result using a self-consistent method such as the one we develop here.

\section{Conclusion}
\label{sec:conclusion}

In this work, we investigate the gas-phase mass-metallicity relation (MZR) and fundamental metallicity relation (FZR) for both non-AGN and AGN-host galaxies by combining IFS data from the MaNGA survey with the Bayesian analysis code NebulaBayes. By using IFS data, we mitigate aperture effects and are able to achieve better AGN-host galaxy light separation than would be possible in unresolved data, while Bayesian methods enable us to mitigate the dependence on nuisance parameters (e.g. ionisation parameters), and enable us to measure oxygen abundance in regions where both stellar and AGN sources contribute ionising light, while treating galaxies that both do and do not host AGN on equal footing. This approach therefore allows use to compare the MZR and FZR of AGN- and non-AGN-host galaxies with minimal bias.

Our main findings are as follows: 
\begin{enumerate}
\item For the galaxies in our sample that show no evidence of AGN activity, we recover MZR and FZR trends consistent with literature values. In particular, our MZR is steeply rising for stellar masses below $M_* \sim 10^{10.5} \Msun$ and flattens at higher masses, and we find that along this sequence oxygen abundance is anti-correlated with star formation rate (SFR). An FZR in which the oxygen abundance varies with the composite parameter $\mu = \log M_* - 0.33 \log\mathrm{SFR}$ is tighter than a simple MZR by an amount that is highly statistically significant.
\item AGN-host galaxies, which we divide into four groups using BPT diagrams -- Composite, LINER, Seyfert, and Ambiguous -- generally have similar oxygen abundances to non-AGN galaxies at stellar masses above $10^{10.5} \Msun$, but are biased to higher oxygen abundance below this stellar mass. The offset between the AGN- and non-AGN MZRs reaches a maximum of $\approx 0.2$ dex for stellar masses $\approx 10^{9.5} \Msun$ (\autoref{fig:mzr_other}). 

\item By contrast, the FZR of AGN-host galaxies is significantly closer to that of non-AGN hosts (\autoref{fig:fzr}); the maximum offset is reduced by $0.07$ dex. This reduction is due to the systematically lower star formation rates of AGN hosts at fixed stellar mass (\autoref{fig:sfms}, which for the FZR shifts AGN-host galaxies to higher $\mu$ than non-AGN galaxies of equal stellar mass, bringing AGN-hosts closer to the non-AGN sequence. However, we do not see statistically significant scatter reduction \textit{within} the AGN-host galaxy sample in going from the MZR to the FZR, and, consistent with this, we find that internal to the AGN sample there is little correlation between star formation rate and oxygen abundance. Thus while the population of AGN host galaxies as a whole is displaced in both SFR and oxygen abundance compared to non-AGN of equal stellar mass, and in a way that is consistent with the non-AGN FZR, there is little correlation between the displacements in oxygen abundance and SFR for individual AGN host galaxies.

\item This finding suggests a scenario whereby the non-AGN FZR is, as has previously been conjectured, a result of accretion events that simultaneously drive up galaxies' SFR and drive down their oxygen abundance. Among AGN hosts these accretion events could be suppressed, causing them to be displaced to lower SFR and higher oxygen abundance than non-AGN galaxies of equal mass. However, once AGN-host galaxies leave the SFMS by this lack of accretion, the SFR declines for an individual system while its metallicity remains mostly unchanged.
\end{enumerate}

In future work, we intend to extend our Bayesian analysis method by adding additional ionisation sources such as shocks and post-AGB stars, which will allow us to use even larger samples and remove some of the remaining biases in our work. We also intend to extend this work to IFS surveys with higher spectral resolution (e.g. SAMI), which will also be useful for disentangling shock-driven emission from star-forming and AGN-driven emission. We also plan to extend this work to the spatially-resolved MZR and FZR for deeper insights into the physical processes driving the FZR.

\section*{Acknowledgements}

We thank the anonymous referee for a detailed report that helped significantly in improving the presentation of our work. KG is supported by the Australian Research Council through the Discovery Early Career Researcher Award (DECRA) Fellowship (project number DE220100766) funded by the Australian Government. 
KG is supported by the Australian Research Council Centre of Excellence for All Sky Astrophysics in 3 Dimensions (ASTRO~3D), through project number CE170100013. 
MRK acknowledges support from the Australian Research Council through Laureate Fellowship award FL220100020. 
EW acknowledges support by the Australian Research Council Centre of Excellence for All Sky Astrophysics in 3 Dimensions (ASTRO 3D), through project number CE170100013.
We thank Adam Thomas for technical supports. We thank Yifei Jin, Trevor Mendel, Andrew Battisti, and Jianling Tang's comments and ideas.

\section*{Data Availability}

The data underlying this article will be shared on reasonable request with the corresponding author.


\bibliographystyle{mnras}
\bibliography{Li23}



\appendix

\section{The Signal-to-noise Ratio Cut}
\label{app:snr}

In this section, we explore the sensitivity or our results to the SNR cut we adopt for the purpose of masking spaxels in \autoref{sec:sample}. For the analysis presented in the main text, we require only SNR $>$ 3 for H$\alpha$ and H$\beta$. Our motivation for applying cuts to these lines only is that several authors have shown that adopting SNR cuts on metal lines can introduce biases (e.g., by selectively removing high-metallicity galaxies from the sample), which in turn can weaken or even reverse correlations between oxygen abundance and SFR  \citep{baker2023,salim2014,thomas2019}. However, we can nevertheless evaluate the effects of an alternative procedure that does mask spaxels with weak metal lines. As our alternative scheme, we mask spaxels that do not satisfy SNR $>$ 1 for $\mathrm{[NII]}\,\lambda 6583$, $\mathrm{[OIII]}\,\lambda 5007$, $\mathrm{[SII]}\,\lambda\lambda 6717,31$, $\mathrm{H}\alpha$, and 
SNR $>10$ in H$\beta$; the former masking criterion matches that adopted by \citet{sanchez2023} to select spaxels eligible for classification, while the latter criterion is twice as strict as that adopted by \citet{thomas2019}.

\begin{figure}
{\includegraphics[width=\columnwidth]{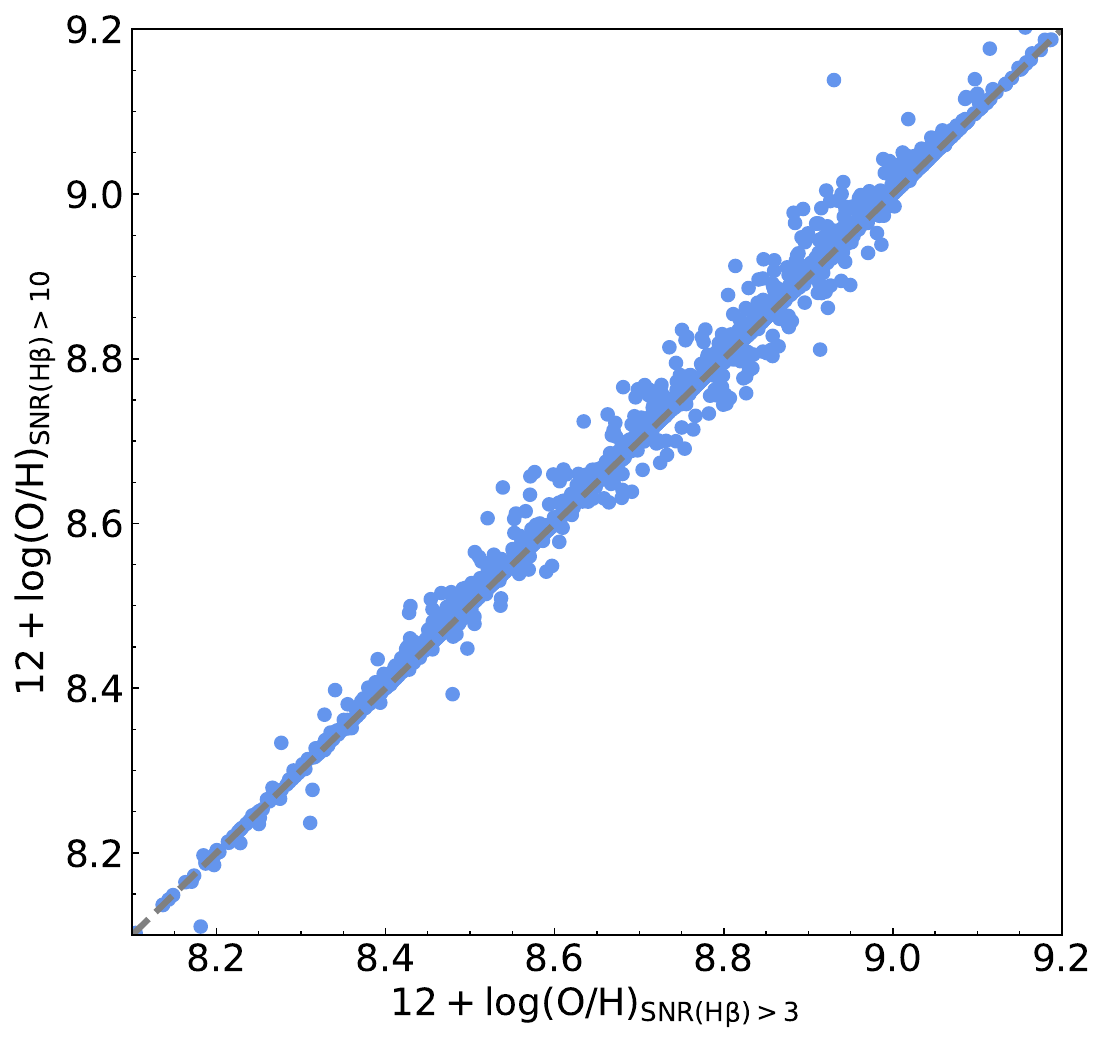}}
    \caption{Comparison of oxygen abundances derived using the SNR criteria adopted in the main text versus the alternative masking scheme described in \aref{app:snr}. The dashed line shows the 1-to-1 relation. The plot shows that the two calculations yield generally consistent results, with a 1$\sigma$ scatter of only $\approx 0.003$ dex between the two methods.}
    \label{fig:snr}
\end{figure}

After applying this alternative masking, re-calculate the oxygen abundances of all galaxies using the same method as in the main text, as described in \autoref{sec:m_sf_abu}. \autoref{fig:snr} compares the oxygen abundances derived using our original masking scheme to those derived using the alternative approach described in this appendix. We see that the results closely follow the 1-to-1 relation, with a 1$\sigma$ scatter of only $\approx 0.003$ dex. This is not surprising, as the mean oxygen abundances we assign to galaxies are H$\alpha$ flux-weighted averages of non-masked spaxels. Spaxels with low SNR are less heavily weighted and therefore have only marginal influence on the final oxygen abundance regardless of how we mask them.

\section{The Setups of the Ionizing Spectra and MAPPINGS}
\label{app:model}

In this appendix, in \autoref{sec:stellar_spectra} and \autoref{sec:agn_spectra} we present details of the libraries of stellar- and AGN-driven ionisation spectra we use to drive our ionisation models, respectively. In \autoref{sec:mappings}, we describe the details of our photoionisation calculations using the MAPPINGS code. 

\subsection{Stellar Ionising Spectra}
\label{sec:stellar_spectra}

To determine our stellar ionising spectra consistent with this abundance scaling, we adopt the Stromlo Stellar Tracks \citep{grasha2021} that focus on massive stars ($10 \le M_*/\Msun \le 300$) and implement GC chemical abundances in their stellar evolution models. The stellar ionising spectra for these models derived using the Flexible Stellar Population Synthesis \citep[FSPS][]{conroy2009,conroy2010} code with the following settings:

\begin{itemize}
    \item Stromlo Stellar Tracks with the rotation $v/v_\mathrm{crit} = 0.4$.
    
    \item A Chabrier IMF \citep{chabrier2003}.
    
    \item A constant SFR of 0.001 $\Msunyr$.
    
    \item A single snapshot of the spectrum after 5 Myr.
    
    \item The BaSeL stellar library from \cite{westera2003} and the carbon star library from \cite{aringer2009}; we interpolate the stellar spectra provided in these library in $\log L - \log Z$ space to match up with our stellar tracks, which are tabulated for a range of $Z/Z_0$. 
\end{itemize}

Using this method we calculate stellar ionising spectra at 9 stellar metallicities: 0.08, 0.20, 0.40, 0.70, 1.00, 1.30, 1.70, 2.00, and 2.50 $Z/Z_0$. This is the same set of metallicities as used by \cite{thomas2018a} except for 0.02 $Z/Z_0$ as it is too low for nearby SF galaxies.

\subsection{AGN Ionising Spectra}
\label{sec:agn_spectra}

We derive AGN ionising spectra for NLRs using OXAF \citep{thomas2016}. This model depends on three parameters that directly describe spectral shape: the energy peak of the emission from the accretion disk Big Blue Bump component $E_\mathrm{peak}$, the power law index of the non-thermal component $\Gamma$, and the proportion of the total luminosity from the non-thermal component $p_\mathrm{NT}$. This direct parameterisation of the spectrum avoids dealing with the degeneracy between black hole mass and AGN luminosity. In this work, following the setup in \cite{thomas2018a}, we fix $\Gamma$ = 2.0 and $p_\mathrm{NT}$ = 0.15 and only vary $\log(E_\mathrm{peak}/\mathrm{keV})$; we use a grid of models whereby this parameter takes on values $-2.0$, $-1.8$, $-1.6$, $-1.4$, $-1.2$, and $-1.0$. We adopt this approach in order to reduce the number of parameter dimensions.

\subsection{MAPPINGS}
\label{sec:mappings}

For each driving ionising spectrum, we use MAPPINGS to calculate the corresponding photoionised line emission. MAPPINGS is a photoionisation and shock modeling code first developed by \cite{binette1985}, and subsequently expanded and improved as described in a number of publications \citep{sutherland1993,dopita2000,groves2004,dopita2013,nicholls2012,sutherland2017}. We use MAPPINGS-V, the latest generation of the code, to derive emission line grids for HII regions and NLRs in this work. Our MAPPINGS setup uses the following parameter choices:

\begin{itemize}
    \item Depletion of elements onto dust grains without the destruction scenario. The fiducial depletion factor for iron is $\log(\mathrm{Fe_{free}/Fe_{total}}) = -1.5$ at $\Oabu = 8.76$, adopted from \cite{jenkins2009,jenkins2014}.
    \item Gas-phase oxygen abundances that take on 11 distinct values, $\Oabu =$ 7.750, 8.150, 8.427, 8.632, 8.760, 8.850, 8.943, 8.997, 9.096, 9.180, and 9.252, corresponding to the metallicities $Z/Z_0 =$ 0.08, 0.20, 0.40, 0.70, 1.00, 1.30, 1.70, 2.00, 2.50, 3.00, and 3.50 used in our stellar ionising spectra. We match the stellar ionising spectra with the gas-phase oxygen abundance in $Z/Z_0$, i.e., our model nebular abundances are always consistent with the stellar abundances we assume when deriving the ionising spectrum. The only exceptions to this statement are $\Oabu =$ 9.180 and 9.252, corresponding to $Z/Z_0 = 3.00$ and $3.50$, for which we assign stellar spectra with $Z/Z_0 =$ 2.5. This is the largest metallicity value that is available in the Stromlo Stellar Tracks. 
    \item To set the abundances of other elements, we adopt the Galactic Concordance (GC) chemical abundance scale \citep{nicholls2017}, which provides a more physically realistic parameterisation for how the relative abundances of elements change with the overall metal abundance than the more commonly-adopted assumption of scaled Solar abundances, which amounts to assuming that relative abundances are invariant. For this purpose we take the ``Solar'' abundance to be $Z_0 = 0.01425$, corresponding to $\Oabu = 8.76$, which is the average derived from high-resolution spectra of 29 local B stars -- see \citet{nieva2012} and \cite{nicholls2017} for details -- and set the abundances of other elements following the prescription provided by \citeauthor{nicholls2017}. In particular, for nitrogen we adopt 
    \begin{equation}
     \log \mathrm{(N/O)}=\log\left[10^{-1.732}+10^{\log\mathrm{(O/H)}+2.19}\right],
    \label{equ:no_ratio}
    \end{equation}
    which fits well for both the primary and secondary nitrogen.
    \item We run MAPPINGS models for 8 values of the gas pressure, $\log(P/k) =$ 4.0, 5.4, 5.8, 6.2, 6.6, 7.0, 7.4, and 8.0 and 9 values of the ionisation parameter at the inner edge of the nebula, with $ -4.00 \le \log U_\mathrm{HII} \le -2.00$ with a constant interval of 0.25. 
    \item All our models assume spherical geometry for the nebula.
\end{itemize}

Our setup for the NLR models is the same as for HII regions in our choice of dust depletion prescription, gas pressure, and oxygen abundance. For the NLR models we consider ionisation parameters at the inner edge for NLRs in the range $ -3.80 \le \log U_\mathrm{NLR} \le -0.20$ with a steps of 0.4, and 6 values of $\log(E_\mathrm{peak}/\mathrm{keV})$ from $-2.00$ to $-1.00$ with steps of 0.2. Finally, we adopt plane parallel geometry for NLR models. The output emission line fluxes of both HII regions and NLRs are all normalised to the H$\beta$ line.

To generate the combined grids including both the HII-region grids and the NLR grids, we first re-normalize all emission lines to H$\alpha$ in both HII-region grids and NLR grids. We then match the HII region and NLR models with the same gas pressure and oxygen abundance, and sum the line fluxes predicted by those models with weights parameterised by the relative contributions of HII region and the NLR emission, $f_\mathrm{HII}$ and $1 - f_\mathrm{HII}$, respectively, where the $f_\mathrm{HII}$ is in the range of $0 \le f_\mathrm{HII} \le 1$ with steps of 0.1. The final combined grids have six dimensions: $\Oabu$, $\log(P/k)$, $\log U_\mathrm{NLR}$, $\log U_\mathrm{HII}$, $\log(E_\mathrm{peak}/\mathrm{keV})$, and $f_\mathrm{HII}$. 

As described in \autoref{sec:NB}, for SF spaxels we only use the part of this grid with $f_\mathrm{HII} = 1$, which leaves only three free parameters: $\log U_\mathrm{HII}$, $\log(P/k)$, and $\Oabu$. For inference in these three parameters we linearly interpolate our MAPPINGS models to a $40\times 20 \times 160$ grid with uniform spacing, which provides a reasonable balance between computation time and the accuracy of the estimation. When fitting non-SF spaxels, we use the full grid and fix the value of $\log U_\mathrm{HII}$ and $\log(E_\mathrm{peak}/\mathrm{keV})$ as discussed in \autoref{sec:NB}. This leaves four free parameters, $\log U_\mathrm{NLR}$, $\log(P/k)$, $\Oabu$, and $f_\mathrm{HII}$, which we linearly interpolate onto a $40\times 20\times  160\times 40$ grid with uniform spacing as input to our NebulaBayes estimation.


\bsp	
\label{lastpage}
\end{document}